\documentclass[]{aa}

\usepackage{graphicx}
\usepackage{txfonts}
\usepackage{float}
\usepackage{booktabs}
\usepackage{multirow}
\usepackage{natbib}
\usepackage{wasysym} 
\bibpunct{(}{)}{;}{a}{}{,}

\usepackage{natbib,twoopt}
\bibpunct{(}{)}{;}{a}{}{,}             
\makeatletter
  \newcommandtwoopt{\citeads}[3][][]{\href{http://adsabs.harvard.edu/abs/#3}%
    {\def\hyper@linkstart##1##2{}%
     \let\hyper@linkend\@empty\citealp[#1][#2]{#3}}}
  \newcommandtwoopt{\citepads}[3][][]{\href{http://adsabs.harvard.edu/abs/#3}%
    {\def\hyper@linkstart##1##2{}%
     \let\hyper@linkend\@empty\citep[#1][#2]{#3}}}
  \newcommandtwoopt{\citetads}[3][][]{\href{http://adsabs.harvard.edu/abs/#3}%
    {\def\hyper@linkstart##1##2{}%
     \let\hyper@linkend\@empty\citet[#1][#2]{#3}}}
  \newcommandtwoopt{\citeyearads}[3][][]%
    {\href{http://adsabs.harvard.edu/abs/#3}
    {\def\hyper@linkstart##1##2{}%
     \let\hyper@linkend\@empty\citeyear[#1][#2]{#3}}}
\makeatother

\newcommand{\teff}{\mbox{$T_{\rm eff}$}}
\newcommand{\logg}{\mbox{$\log g$}}
\newcommand{\vsini}{\mbox{$v \sin i$}}
\newcommand{\teq}{\mbox{$T_{\rm eq}$}}
\newcommand{\mictrb}{\mbox{$\xi_{\rm t}$}}
\newcommand{\mactrb}{\mbox{$v_{\rm mac}$}}

\newcommand{\kms}{\mbox{km\,s$^{-1}$}}
\newcommand{\ms}{\mbox{m\,s$^{-1}$}}

\newcommand{\gammaS}{\mbox{$\gamma_{\rm \tiny SOPHIE}$}}
\newcommand{\gammaC}{\mbox{$\gamma_{\rm \tiny CORALIE}$}}

\newcommand{\rhostar}{\ensuremath{\rho_\star}}
\newcommand{\rhosun}{\ensuremath{\rho_\odot}}
\newcommand{\rhoj}{\ensuremath{\rho_{\rm J}}}
\newcommand{\rhopl}{\ensuremath{\rho_{\rm pl}}}
\newcommand{\rj}{R\ensuremath{_{\rm J}}}
\newcommand{\mj}{M\ensuremath{_{\rm J}}}
\newcommand{\me}{M\ensuremath{_\oplus}}
\newcommand{\re}{R\ensuremath{_\oplus}}
\newcommand{\rsun}{R\ensuremath{_\odot}}
\newcommand{\msun}{M\ensuremath{_\odot}}
\newcommand{\rpl}{\ensuremath{R_{\rm pl}}}
\newcommand{\mpl}{\ensuremath{M_{\rm pl}}}
\newcommand{\rstar}{\ensuremath{R_\star}}
\newcommand{\mstar}{\ensuremath{M_\star}}

\def\secos{$\sqrt{e} \cos \omega$}
\def\sesin{$\sqrt{e} \sin \omega$}
\def\feh{[Fe/H]}

\begin{document}

   \title{WASP-86b and WASP-102b: super-dense versus bloated planets}

\author{
F. Faedi \inst{\ref{warwick}}
  \and Y. G\'omez Maqueo Chew \inst{\ref{unam}}
  \and D. Pollacco \inst{\ref{warwick}} 
  \and D. J. A. Brown \inst{\ref{warwick}} 
  \and G. H\'ebrard \inst{\ref{iap},\ref{ohp} } 
  \and B. Smalley \inst{\ref{keele}} 
  \and K. W. F. Lam \inst{\ref{warwick}}
  \and D. Veras \inst{\ref{warwick}}
  \and D. Anderson \inst{\ref{keele}}
  \and A. P.\ Doyle \inst{\ref{warwick}}
  \and M. Gillon \inst{\ref{liege}}
  \and M. R. Goad \inst{\ref{leic}}
  \and M. Lendl \inst{\ref{austria},\ref{genv}}
  \and L. Mancini \inst{\ref{heid}}  
  \and J. McCormac \inst{\ref{warwick}} 
  \and I. Plauchu-Frayn \inst{\ref{uname}} 
  \and J. Prieto-Arranz \inst{\ref{iac},\ref{ull}} 
  \and A. Scholz \inst{\ref{sta}}
  \and R. Street \inst{\ref{lcogt}}
  \and A. H. M. Triaud \inst{\ref{genv},\ref{cam}}
  \and R. West \inst{\ref{warwick}}
  \and P. J. Wheatley \inst{\ref{warwick}}
  \and D. J. Armstrong\inst{\ref{warwick},\ref{qub}}
  \and S. C. C. Barros \inst{\ref{porto}}
  \and I. Boisse \inst{\ref{lam}}
  \and F. Bouchy \inst{\ref{genv},\ref{lam}}
  \and P. Boumis \inst{\ref{greece}}  
  \and A. Collier Cameron \inst{\ref{sta}}
  \and C. A. Haswell \inst{\ref{ou}}
  \and K. L. Hay \inst{\ref{sta}}
  \and C. Hellier \inst{\ref{keele}}
  \and U. Kolb \inst{\ref{ou}}
  \and P. F. L. Maxted \inst{\ref{keele}}
  \and A. J. Norton \inst{\ref{ou}}
  \and H. P. Osborn \inst{\ref{warwick}}
  \and E. Palle \inst{\ref{iac},\ref{ull}}
  \and F. Pepe \inst{\ref{genv}}
  \and D. Queloz \inst{\ref{cam},\ref{genv}}
  \and D. S\'egransan \inst{\ref{genv}}
  \and S. Udry \inst{\ref{genv}}
  \and P. A. Wilson \inst{\ref{iap}}
}

\institute{Department of Physics, University of Warwick, Coventry CV4
  7AL, UK \email{f.faedi@warwick.ac.uk}\label{warwick} \and Instituto
  de Astronom\'ia, Universidad Nacional Aut\'onoma de M\'exico, Ciudad
  Universitaria, Ciudad de M\'exico, M\'exico\label{unam} \and
  Institut d'Astrophysique de Paris, UMR7095 CNRS, Universit\'e Pierre
  \& Marie Curie, 98bis boulevard Arago, 75014 Paris,
  France\label{iap} \and Observatoire de Haute-Provence, Universit\'e
  d’Aix-Marseille \& CNRS, 04870 Saint Michel l’Observatoire,
  France\label{ohp} \and Astrophysics Group, Keele University,
  Staffordshire, ST5 5BG, UK\label{keele} \and Universit\'e de
  Li\`ege, All\'ee du 6 ao$\hat {\rm u}$t 17, Sart Tilman, Li\`ege 1,
  Belgium\label{liege} \and Department of Physics and Astronomy,
  University of Leicester, Leicester, LE1 7RH, UK\label{leic} \and
  Space Research Institute, Austrian Academy of Sciences,
  Schmiedlstr. 6, 8042 Graz, Austria\label{austria} \and Observatoire
  astronomique de l'Universit\'e de Gen\`eve, 51 ch. des Maillettes,
  1290 Sauverny, Switzerland\label{genv} \and Max Planck Institute for
  Astronomy, Heidelberg, Germany\label{heid} \and Instituto de
  Astronom\'ia, Universidad Nacional Aut\'onoma de M\'exico,
  C.P. 22860, Ensenada, Baja California, M\'exico \label{uname} \and
  Instituto de Astrof\'isica de Canarias (IAC), 38205 La Laguna,
  Tenerife, Spain \label{iac} \and Departamento de Astrof\'isica,
  Universidad de La Laguna (ULL), 38206 La Laguna, Tenerife,
  Spain\label{ull} \and School of Physics and Astronomy, University of
  St Andrews, St Andrews, Fife KY16 9SS, UK \label{sta} \and Las
  Cumbres Observatory Global Telescope Network, 6740 Cortona Drive,
  Suite 102, Santa Barbara,CA 93117, USA\label{lcogt} \and Department
  of Physics, University of Cambridge, J J Thomson Av, Cambridge, CB3
  0HE, UK \label{cam} \and Astrophysics Research Centre, Queen's
  University Belfast, University Road, Belfast BT7 1NN, UK \label{qub}
  \and Instituto de Astrof\'isica e Ci\^{e}ncias do Espa\c co,
  Universidade do Porto, CAUP, Rua das Estrelas, 4150-762 Porto,
  Portugal \label{porto} \and Aix Marseille Universit\'e, CNRS, LAM
  (Laboratoire d'Astrophysique de Marseille) UMR 7326, 13388
  Marseille, France \label{lam} \and Institute for Astronomy,
  Astrophysics, Space Applications and Remote Sensing, National
  Observatory of Athens, 15236 Penteli, Greece \label{greece} \and
  Department of Physical Sciences, The Open University, Milton Keynes,
  MK7 6AA, UK \label{ou} }


\date{Received; accepted}

\abstract{ We report the discovery of two transiting planetary
  systems: a super dense, sub-Jupiter mass planet WASP-86b (\mpl\ =
  0.82 $\pm$ 0.06 \mj; \rpl\ = 0.63 $\pm$ 0.01 \rj), and a bloated,
  Saturn-like planet WASP-102b (\mpl\ = 0.62 $\pm$ 0.04 \mj; \rpl\ =
  1.27 $\pm$ 0.03 \rj). They orbit their host star every $\sim$5.03,
  and $\sim$2.71 days, respectively. The planet hosting WASP-86 is a
  F7 star (\teff\ = 6330$\pm$110 K, \feh\ = $+$0.23 $\pm$ 0.14 dex,
  and age $\sim$0.8--1~Gyr); WASP-102 is a G0 star (\teff\ =
  5940$\pm$140 K, \feh\ = $-$0.09$\pm$ 0.19 dex, and age
  $\sim$1~Gyr). These two systems highlight the diversity of planetary
  radii over similar masses for giant planets with masses between
  Saturn and Jupiter. WASP-102b shows a larger than model-predicted
  radius, indicating that the planet is receiving a strong incident
  flux which contributes to the inflation of its radius. On the other
  hand, with a density of $\rhopl$ = 3.24$\pm$~0.3~$\rhoj$, WASP-86b
  is the densest gas giant planet among planets with masses in the
  range 0.05 $<\mpl<$ 2.0 \mj. With a stellar mass of 1.34 $\msun$ and
  \feh = $+$0.23 dex, WASP-86 could host additional massive and dense
  planets given that its protoplanetary disc is expected to also have
  been enriched with heavy elements.  In order to match WASP-86b's
  density, an extrapolation of theoretical models predicts a planet
  composition of more than 80\% in heavy elements (whether confined in
  a core or mixed in the envelope).  This fraction corresponds to a
  core mass of approximately 210\me\ for WASP-86b's mass of
  \mpl$\sim$260\,\me. Only planets with masses larger than about 2
  \mj\ have larger densities than that of WASP-86b, making it
  exceptional in its mass range.}


\maketitle
%

\section{Introduction}

We now know of more than 2500 planetary systems with single and
multiple planets \footnote{http://exoplanet.eu}.  Among these
discoveries, the WASP sample represents an important contribution
because WASP planets orbit bright stars which allow for precise
follow-up photometric and spectroscopic observations. To date, the
WASP survey \citep{Pollacco2006} has discovered more than 160 planets,
making it the most successful ground-based transit survey. We are now
in the era of K2 \citep{Howell2014}, TESS \citep{Ricker2015}, and
CHEOPS space missions hunting for Earth-like analogues. However,
ground-based wide-fields surveys, such as WASP and HAT/HATS
(\citealt{Bakos2004,Bakos2013}) just to mention a few, are capable of
detecting peculiar objects for example HATS-17 b, \citep{Panev2016};
HATS-18 b, \citep{Brahm2016}, increasing the spectrum of possible
mass-radius relations in the planetary regime. These systems provide
invaluable observational constraints on theoretical models.

Bright transiting systems are the only systems for which masses and
radii can be derived with high precision, in turn providing insight
into the planetary bulk composition via their estimated densities. The
wide range of properties observed for the class of gas giant planets
are still not fully understood. For example, the diversity in
exoplanet densities and hence in their internal compositions is
particularly noticeable at sub-Jupiter masses (0.05 $< \mpl <$ 1\mj)
where densities span 2 orders of magnitude.  Systems like HD\ 149026b
($\rhopl\simeq 1\rhoj$; \citealt{Sato2005}) and WASP-59b
($\rhopl\simeq 1.8\rhoj$; \citealt{Hebrard2013}) are very dense
planets in this mass range for which a rock/ice core of $\sim$70~$\me$
(corresponding to a heavy elements enrichment of $>60\%$) is
hypothesised.  At the opposite end of the spectrum we have planets
like WASP-17b ($\rhopl = 0.06 \rhoj$,
\citealt{Anderson2010a,Anderson2011b}) and WASP-31b ($\rhopl =
0.132~\rhoj$,~\citealt{Anderson2010b}), which are examples of planets
with puzzling low densities.  However, the planets in these systems
are strongly irradiated. One of our latest discoveries is WASP-127b
\citep{Lam2016}; with a mass of $\sim$3 times that of Neptune ($\mpl =
0.18~\mj$) and a radius of 1.3$\rj$ it is another example of an
extremely low density planet ($\rhopl = 0.068 \rhoj$). However its
host star is a G0 and its orbital period is $\sim4$\,d implying that
WASP-127b does not receive the same amount of flux as the two examples
mentioned above.  To assess the inflation status of a system,
generally planetary radii are compared to theoretical models
\citep[e.g.,][]{Fortney2007,Burrows2007,Baraffe2008}.  However, the
radius depends on multiple physical properties such as the stellar
age, the irradiation flux, the planet's mass, the atmospheric
composition, the presence of heavy elements in the envelope or in the
core, the atmospheric circulation, and also on any source generating
extra heating in the planetary interior. Although models account for
these contributions (e.g., tidal heating due to unseen companions
pumping up the eccentricity \citealt{Bodenheimer2001},
\citealt{Bodenheimer2003}; kinetic heating due to the breaking of
atmospheric waves \citealt{Guillot2002}; enhanced atmospheric opacity
\citealt{Burrows2007}; and semi-convection \citealt{Chabrier2007}),
they can not explain the entire range of observed radii
\citep{Fortney2010,Leconte2010}.  This is not just the case for
Jupiter-like gas giant planets, as even Neptune-like and smaller
super-Earth planets show a large variety of properties which are
difficult to reconcile with current knowledge of internal composition,
structure, and formation histories (see for example
\citealt{Lissauer2014,Mayor2014}).\\

In this paper, we present the discovery of two new transiting
planetary systems from the WASP Survey: 1SWASP J175033.71+363412.7,
hereafter WASP-86, and 1SWASP J222551.44+155124.5, hereafter WASP-102.
WASP-86b and WASP-102b belong to the class of gas giant planets with
sub-Jupiter masses. WASP-86b is the densest gas giant planet with a
mass between that of Neptune and twice Jupiter's mass, and shows
similarities to both WASP-59b and HATS-17b.  WASP-102b, in contrast,
is a very bloated planet with a mass twice that of Saturn showing a
radius anomaly similarly to WASP-17b and WASP-12b. Thus these
SuperWASP discoveries provide new evidence of more extreme systems.

The paper is structured as follows: in $\S \ref{obs}$ and $\S
\ref{rvobs}$ we describe the observations, including the WASP
discovery data and follow-up photometric and spectroscopic
observations which establish the planetary nature of the transiting
objects. In \S \ref{properties} we present our results for the derived
system parameters for the two systems, as well as the individual
stellar and planetary properties. Finally in \S \ref{discussion}, we
discuss the implication of these discoveries, their physical
properties and how they extend the currently known mass-radius
parameter space.

\section{Photometric Observations}\label{obs}

\subsection{WASP Photometry}\label{swobs}

The SuperWASP telescope is located at the Roque de los Muchachos
Observatory in La Palma (ING, Canary Islands, Spain).  The telescope
consists of 8 Canon 200mm f/1.8 lenses coupled to e2v 2048$\times$2048
pixel CCDs, which yield a field of view of $7.8\times7.8$ square
degrees with a corresponding pixel scale of 13\farcs7
\citep{Pollacco2006}.  The SuperWASP observations have exposure times
of 30 seconds, and a typical cadence of 8 min during the observing
season.  All WASP data are processed by the custom-built reduction
pipeline described in \citet{Pollacco2006}.  The resulting light
curves are analysed using our implementation of the Box Least-Squares
and SysRem detrending algorithms \citep[see
][]{Cameron2006,Kovacs2002,Tamuz2005} to search for transit-like
features.  Once the targets are identified as planet candidates a
series of multi-season, multi-camera analyses are performed on the
WASP photometry to strengthen the candidate's detection.  These
additional tests allow a more thorough analysis of the stellar and
planetary parameters, which are derived solely from the WASP data and
publicly available catalogues \citep[e.g., UCAC4;][]{Zacharias2013},
thus helping in the identification of the best candidates, as well as
the rejection of possible spurious detections.\\

\begin{figure}[!ht]
\centering
\includegraphics[width=0.5\textwidth]{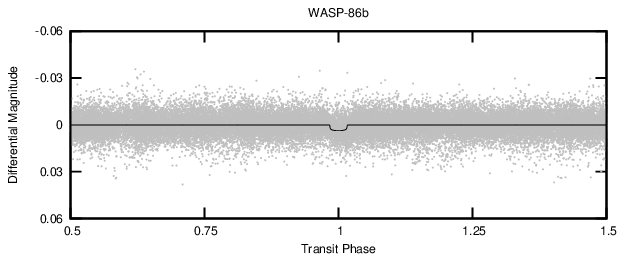}
\includegraphics[width=0.5\textwidth]{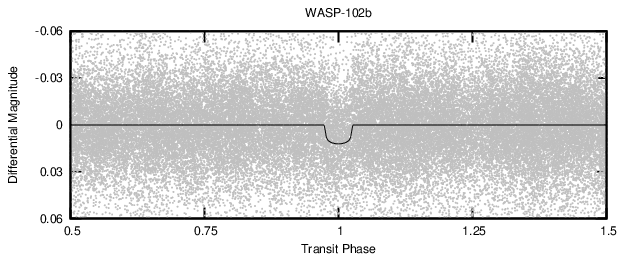}
 \caption{ Discovery WASP Light Curves. 
 {\em Upper panel}: WASP transit light curve of WASP-86b, 
phase folded on the ephemeris given in Table~\ref{planets_params}. 
The black, solid line is the MCMC best-fit transit model, as described in \S\ref{mcmc}. 
{\em Lower panel}: Same as above in the case of WASP-102b.  
\label{swlcs}
}
\end{figure}

\begin{figure}[!ht]
\centering
\includegraphics[width=0.5\textwidth]{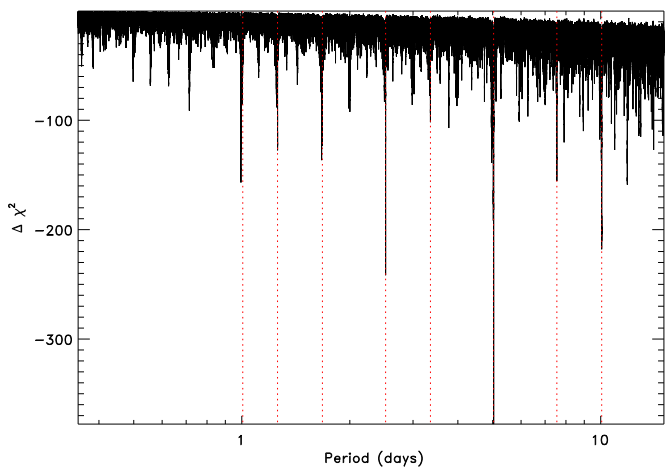} 
 \caption{ Box-Least Square periodogram of WASP-86b lightcurve.
Period from the strongest feature is 5.03157 days and from the 
first MCMC run with only SuperWASP data is P = 5.03160 $\pm$ 0.00002 days.
The orbital period of WASP-86b derived in \S \ref{mcmc} is the strongest feature in the periodogram;
its aliases are marked by dotted-red lines (which from left to right are: P/5, P/4,P/3,P/2, 2P/3, 3P/2 and 2P).
\label{periodo}
}
\end{figure}

In the case of WASP-86, the WASP-North light curve consists of a total
of 40223 data points that span from 2004 May 03 to 2010 August 24 (see
top panel of Fig.~\ref{swlcs}).  The WASP data show a dip in
brightness characteristic of a transiting planet signal with a period
of $P=5.031$~days, a transit duration of $\sim$4 hours, and a very
shallow transit depth of 2.8~mmag.  Given the very small signal of
WASP-86b its detection is at the limit of the SuperWASP detection
capability.  Although the median photometric error of the WASP
observations is of the same order of magnitude as the dip in
brightness due to the planet, because of the multi-year span of the
light curve and the large number of data points the transit signal is
clearly significant in the periodogram (see Fig.~\ref{periodo}).  The
periodogram is the result of the WASP analysis pipeline in which the
Box-Least Squared periodogram is computed as per the prescription in
\citet{Cameron2006}, and has been modified to fit multiple box widths.
Thus, the reported $\Delta \chi^2$ is of the best epoch and best box-width combination.\\

The WASP-102 WASP-North light curve is comprised of 50126 photometric
measurements spanning from 2004 June 23 to 2011 November 10 (see
bottom Fig.~\ref{swlcs}).  The transit signal in the WASP light curve
is clearly detected, is periodic with a period $P\sim$2.71~d, and has
a width of $\sim$3.5 hours and a depth of 10~mmag.

We omit the periodogram of WASP-102b in the paper, because the
detection is more robust and has better follow-up photometry than
WASP-86b which because of a much smaller radius and near-integer
period as well as long transit duration has been elusive to acquire.

\begin{table}[!ht] 
\caption[]{Photometric and astrometric properties of 
    WASP-86 and WASP-102 from UCAC4 \citep{Zacharias2013}.}
\label{table1}
\begin{center}
\begin{tabular}{lccc}
\hline
\hline \\
 Parameter    & WASP-86 && WASP-102  \\
 \hline \\
${\rm RA (J2000)}$		& 17:50:33.718  && 22:25:51.447\\	
${\rm Dec (J2000)}$		& $+$36:34:12.79&&$+$15:51:24.29\\	
${\rm B}$			&$11.33\pm0.06$	&&$13.36\pm0.03$ \\
${\rm V}$			&$10.66\pm0.05$	&&$12.73\pm0.02$\\
${\rm g}$			&$...$		&&$13.06\pm0.17$\\	
${\rm r}$			&$...$		&&$12.55\pm0.04$\\	
${\rm i}$			&$...$ 		&&$13.11\pm0.90$\\	
${\rm J}$			&$9.63\pm0.02$ 	&&$11.49\pm0.02$\\
${\rm H}$			&$9.39\pm0.02$ 	&&$11.22\pm0.02$\\	
${\rm K}$			&$9.36\pm0.02$ 	&&$11.11\pm0.02$\\	
$\mu_{\alpha}$ (mas/yr)	        &$-1.7\pm1.0$	&&$-12.0\pm2.0$	\\	
$\mu_{\delta}$   (mas/yr)	&$-9.7\pm0.7$	&&$-22.6\pm2.6$\\	
\hline\\
\end{tabular}
\end{center}
\end{table}


\subsection{Follow-Up Multi-band Photometry}\label{fuobs}

In this section, we describe the follow-up photometric observations
for both systems. We note that given the long transit duration and the
near integer orbital period a full transit light curve of WASP-86b has
been difficult to acquire.
All light curves will be available in the online version of the paper as electronic tables. \\

\begin{table}
\caption{Log of follow-up transit photometry observations.} 
\begin{center}
\begin{tabular}{cccc} 
\hline \hline\\
Planet & Date & Tel./Inst. & Filter \\ 
\midrule
{\multirow{2}{*}{WASP-86b}}	
                & 2013 07 16 & FTN   & Pan-STARRS-Z \\ 
				& 2014 04 23 & NITES & no filter \\ 
                & 2014 08 27 & LT    & V$+$R \\ 
				& 2015 04 15 & SPM   & Johnson R \\ 
\midrule
{\multirow{6}{*}{WASP-102b}}
				& 2013 08 05 & NITES & no filter \\ 
				& 2013 08 13 & TRAPPIST & {\it blue-blocking} \\
				& 2013 08 13 & EulerCam & Gunn-$r$ \\
				& 2013 09 20 & TRAPPIST & {\it blue-blocking}\\
                & 2013 09 20 & EulerCam & Gunn-$r$ \\
				& 2013 10 09 & TRAPPIST & {\it blue-blocking}\\
\bottomrule
\end{tabular}
\label{tablephot}
\end{center}
\end{table} 

\begin{figure*}
  \centering
   \includegraphics[width=0.9\textwidth]{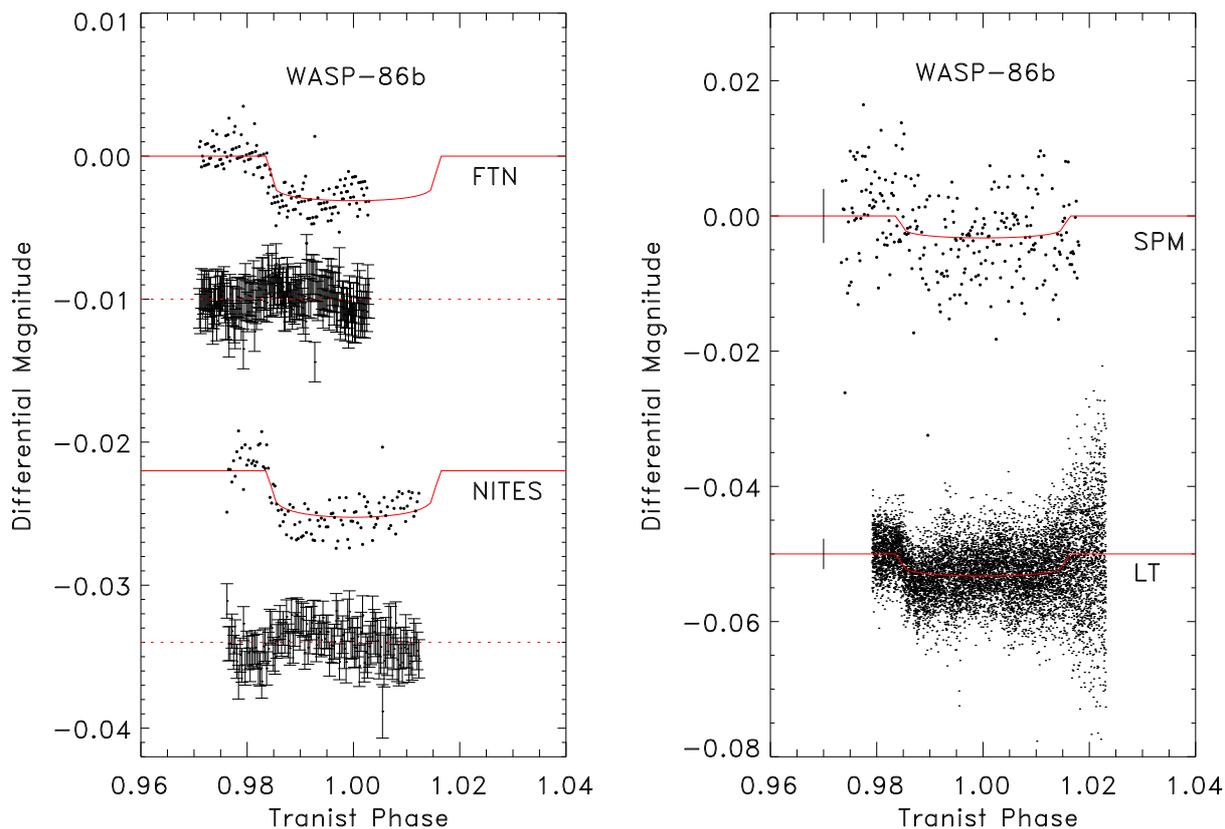} 
   \caption{ Follow-up, high-precision, time-series photometry of
     WASP-86b during transit (see Table~\ref{tablephot}).  The
     observations are shown as black points and are phase folded on
     the ephemeris shown in Table~\ref{planets_params}.  The
     superimposed, solid, red line is our best-fit transit model
     (\S\ref{mcmc}) using the formalism of \citet{Mandel2002}. The
     residuals from the fit and the individual data points photometric
     uncertainties are displayed directly under each light
     curve. Light curves and residuals are displaced from zero for
     clarity. In the right panel (SPM and LT light curves) given the
     density of the data and large uncertainties, we plot the average
     errorbar on the left of each light curve to avoid crowding.}
     \label{W86_lcs}
    \end{figure*}

\begin{figure}
   \centering
   \includegraphics[width=0.5\textwidth]{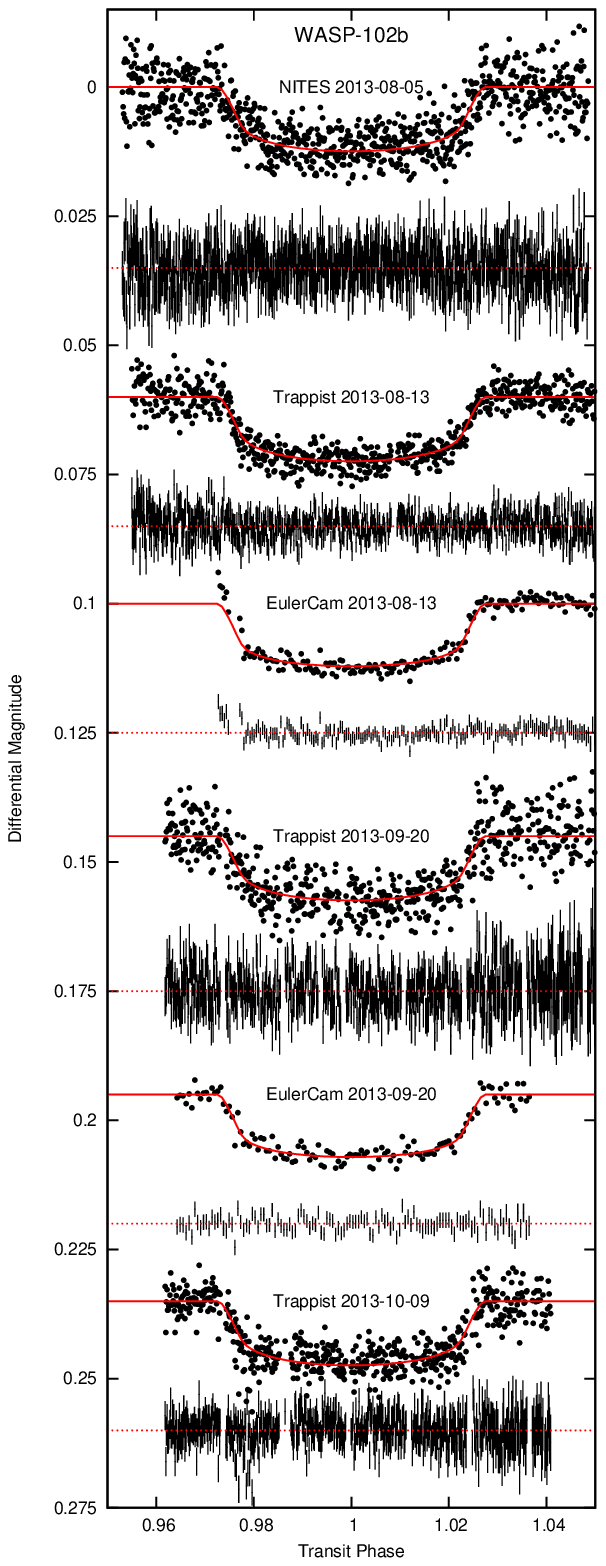}
   \caption{ Follow-up, high signal-to-noise light curves of WASP-102b
     during transit.  Same as Figure~\ref{W86_lcs}.  }
     \label{W102_lcs}
    \end{figure}

{\bf Faulkes North Telescope Observations.}
WASP-86 was observed with the 2-m Faulkes North Telescope in Hawai'i,
USA on 2013 July 16, using the Spectral camera with a Pan-STARRS-Z
filter.  This has a Fairchild CCD486 BI detector with a pixel scale of
0.304 arcsec pixel$^{-1}$ in the (default) 2$\times$2 binning
mode. The instrument was deliberately defocussed in order to spread
the light over a larger number of pixels and to avoid saturation while
executing 60s exposures, long enough to minimise noise due to
scintillation. The data were pre-processed using the standard LCOGT
2-m reduction pipeline in use at the time, since these data were
acquired prior to the 2-m telescopes being fully integrated with the
larger LCOGT network. Aperture photometry was
then conducted using a stand-alone implementation of DAOPHOT \citet{Stetson1987}.\\

{\bf NITES Observations.}
The Near Infra-red Transiting ExoplanetS (NITES) Telescope is a
semi-robotic $0.4$-m (f/10) Meade LX200GPS Schmidt-Cassegrain
telescope installed at the Observatorio del Roque de los Muchachos, La
Palma, Spain. The telescope is mounted with Finger Lakes
Instrumentation Proline 4710 camera, containing a $1024\times1024$
pixels deep-depleted CCD made by e2v. The telescope has a field of
view of $11\times11$ arcmin squared and a pixel scale of 0.66 arcsec
pixel$^{-1}$, respectively, and a peak QE$>90\%$ at $800$ nm. For more
details on the NITES Telescope we refer the reader to
\citet{McCormac2014}.

One transit of WASP-86 b was observed on 2014 April 23. The telescope
was defocused slightly to 7.3 arcsec (FWHM) and $587$ images of $20$ s
exposure time were obtained with $5$ s dead time between each. The
dead time is a combination of the CCD readout and an additional dwell
time to allow for science frame autoguiding using the DONUTS algorithm
\citep{McCormac2013}.
One transit of WASP-102 b was observed on 2013 August 05. The
telescope was defocused slightly to 3.3 arcsec and $827$ images of
$20$ s exposure time were obtained with $5$ s dead time between each.

In order to obtain the best signal-noise ratio (SNR) both observations
were made without a filter. The data were bias subtracted and
flat-field corrected using PyRAF\footnotemark \footnotetext{PyRAF is a
  product of the Space Telescope Science Institute, which is operated
  by AURA for NASA.}  and the standard routines in IRAF\footnotemark
\footnotetext{IRAF is distributed by the National Optical Astronomy
  Observatories, which are operated by the Association of Universities
  for Research in Astronomy, Inc., under cooperative agreement with
  the National Science Foundation.} and aperture photometry was
performed using DAOPHOT \citep{Stetson1987}. A total of $5$ and $6$
nearby comparison stars were used and aperture radii of $12\arcsec$
and $4\arcsec$ were chosen as they returned the minimum RMS scatter in
the out of transit data for WASP-86 b and WASP-102 b,
respectively. Initial photometric error estimates were calculated
using the electron noise from the
target and the sky, and the read noise within the aperture.\\

{\bf San Pedro M\'artir Observations.}
The transit of WASP-86b was observed with the 84cm Telescope at the
Observatorio Astron\'omico Nacional de San Pedro M\'artir (SPM) in
Baja California, M\'exico on 2015 April 15 (UT) using the Marconi 3
CCD and MEXMAN filter wheel.  The images were binned 2$\times$2. The
telescope was defocussed such that the exposure times were 60 s to
maximise time on target and minimise the effects of the shutter, read
out and scintillation.  A total of 258 photometric data points using
the Johnson R filter were acquired.  We also observed the WASP-86b
transit on 2015 April 11 with the same configuration; however due to
clouds and the shallowness of the WASP-86 transits, the data were not
sufficiently good to include in the analysis.
The data were reduced and the light curves extracted following the 
standard procedures described in \S\ref{fuobs}: NITES Observations.\\

{\bf TRAPPIST Observations.}
Three transits of WASP-102b were observed with the 0.6-m TRAPPIST
robotic telescope (TRAnsiting Planets and PlanetesImals Small
Telescope), located at ESO La Silla Observatory, Chile. TRAPPIST is
equipped with a thermoelectrically-cooled 2K$\times$2K CCD, which has
a pixel scale of 0.65\arcsec\ that translates into a
22\arcmin$\times$22\arcmin\ field of view. For details of TRAPPIST,
see \citet{Gillon2011} and \citet{Jehin2011}. The TRAPPIST photometry
was obtained using a readout mode of 2$\times$2 MHz with 1$\times$1
binning, resulting in a readout plus overhead time of 6.1~s and a
readout noise of 13.5 e$^{-}$. A slight defocus was applied to the
telescope to improve the duty cycle, spread the light over more
pixels, and, thereby improve the sampling of the point-spread function
(PSF). The transits were observed in a blue-blocking
filter\footnote{http://www.astrodon.com/products/filters/exoplanet/}
that has a transmittance $> 90\%$ from 500~nm to beyond 1000~nm.
During the runs, the positions of the stars on the chip were
maintained to within a few pixels thanks to a ''software guiding''
system that regularly derives an astrometric solution for the most
recently acquired image and sends pointing corrections to the mount if
needed. After a standard reduction (bias, dark, and flat-field
correction), the stellar fluxes were extracted from the images using
the IRAF/ DAOPHOT aperture photometry software \citep{Stetson1987}.
For each light curve we tested several sets of reduction parameters
and kept the one giving the most precise photometry for the stars of
similar brightness as the target. After a careful selection of
reference stars, the transit light curves were finally obtained using
differential photometry.\\

{\bf EulerCam Observations.}
We observed two transits of WASP-102 using EulerCam mounted on the
1.2-m Swiss Telescope at ESO La Silla, Chile \citep{Lendl2012}. On
2013 August 13, we used an Gunn-r$^\prime$ filter and 80 s exposure
times, obtaining 182 images with stellar FWHM between 1.15\arcsec\ and
1.9\arcsec; while on 2013 September 20, we observed through a
Cousins-I filter obtaining 123 120s exposures with FWHM between
1.9$\arcsec$ and 5.05$\arcsec$. All data were reduced as outlined in
\citet{Lendl2012}, and we performed aperture photometry with apertures
of 5.05\arcsec (2013 Aug 13) and 9.45\arcsec (2013 Sep 20), for the
first and second transit light curve, respectively.  We carefully
selected the most stable field stars as reference stars for the
relative differential photometry such that the scatter in the light
curves was minimised.

\section{Spectroscopic Observations} \label{rvobs}

WASP-86 and -102 were observed during our follow-up campaigns between
2012 May 16 and 2015 November 8 by means of the SOPHIE spectrograph
mounted at the 1.93-m telescope (\citealt{Perruchot2008};
\citealt{Bouchy2009}) at Observatoire de Haute-Provence (OHP). In
addition, WASP-102 was observed between 2012 September 15 and 2014
August 19 with the CORALIE spectrograph mounted at the 1.2-m Euler
Swiss telescope at La Silla, Chile (\citealt{Baranne1996};
\citealt{Queloz2000}; \citealt{Pepe2002}).

The SOPHIE observations were obtained in high efficiency mode (R =
40\,000), with very similar signal-to-noise ratio ($\sim$30), in order
to minimise systematic errors (e.g., the charge transfer inefficiency
effect of the CCD, \citealt{Bouchy2009}). Wavelength calibration with
a thorium-argon lamp was performed every $\sim$2 hours, allowing for
interpolation of the spectral drift of SOPHIE ($<$3 \ms per hour; see
\citealt{Boisse2010}). Two 3$\arcsec$ diameter optical fibers were
used; the first centered on the target and the second on the sky to
simultaneously measure the background to remove contamination from
scattered moonlight.  The contamination of the CCF from scattered
moonlight was negligible for most of the SOPHIE exposures because the
Moon was low and/or well shifted from the targets' radial velocity
(RV) measurements. In some cases however there was a significant
contamination which was corrected using the second SOPHIE
aperture. The maximum corrections were 250 \ms\ for WASP-86, and 160
\ms\ for WASP-102.

The CORALIE observations of WASP-102 were obtained during grey/dark
time to minimise moonlight contamination. Both SOPHIE and CORALIE
data-sets were processed with standard data reduction pipelines.  The
radial velocity uncertainties were evaluated including known
systematics such as guiding and centering errors \citep{Boisse2010},
and wavelength calibration uncertainties. All spectra were
single-lined.  In addition to the radial velocity variation of
WASP-86b, the SOPHIE data show a linear drift which we have been
monitoring. With the current phase coverage of SOPHIE data we can not
yet put constraints on the mass of the secondary object which could be
in the planetary regime.

For each planetary system the radial velocities were computed from a
weighted cross-correlation of each spectrum with a numerical mask of
spectral type G2, as described in \citet{Baranne1996} and
\citet{Pepe2002}. To test for possible stellar impostors we performed
the cross-correlation with masks of different stellar spectral types
(e.g., F0, K0 and K5). For each mask, we obtained similar radial
velocity variations, thus rejecting a blended eclipsing system of
stars with unequal masses as a possible cause of the variation.

We present in Tables \ref{WASP-86_rvtable} and \ref{WASP-102_rvtable}
the spectroscopic measurements of WASP-86 and 102.  In each table we
list the Barycentric Julian date (BJD-TDB), the stellar radial
velocity measurements, their uncertainties, the bisector span
measurements (V$_{\rm span}$), and the residuals to the best-fit
Keplerian model; additionally in the case of WASP-102, we list the
instrument used.
 
\begin{figure*}
   \centering
   \includegraphics[width=0.49\textwidth]{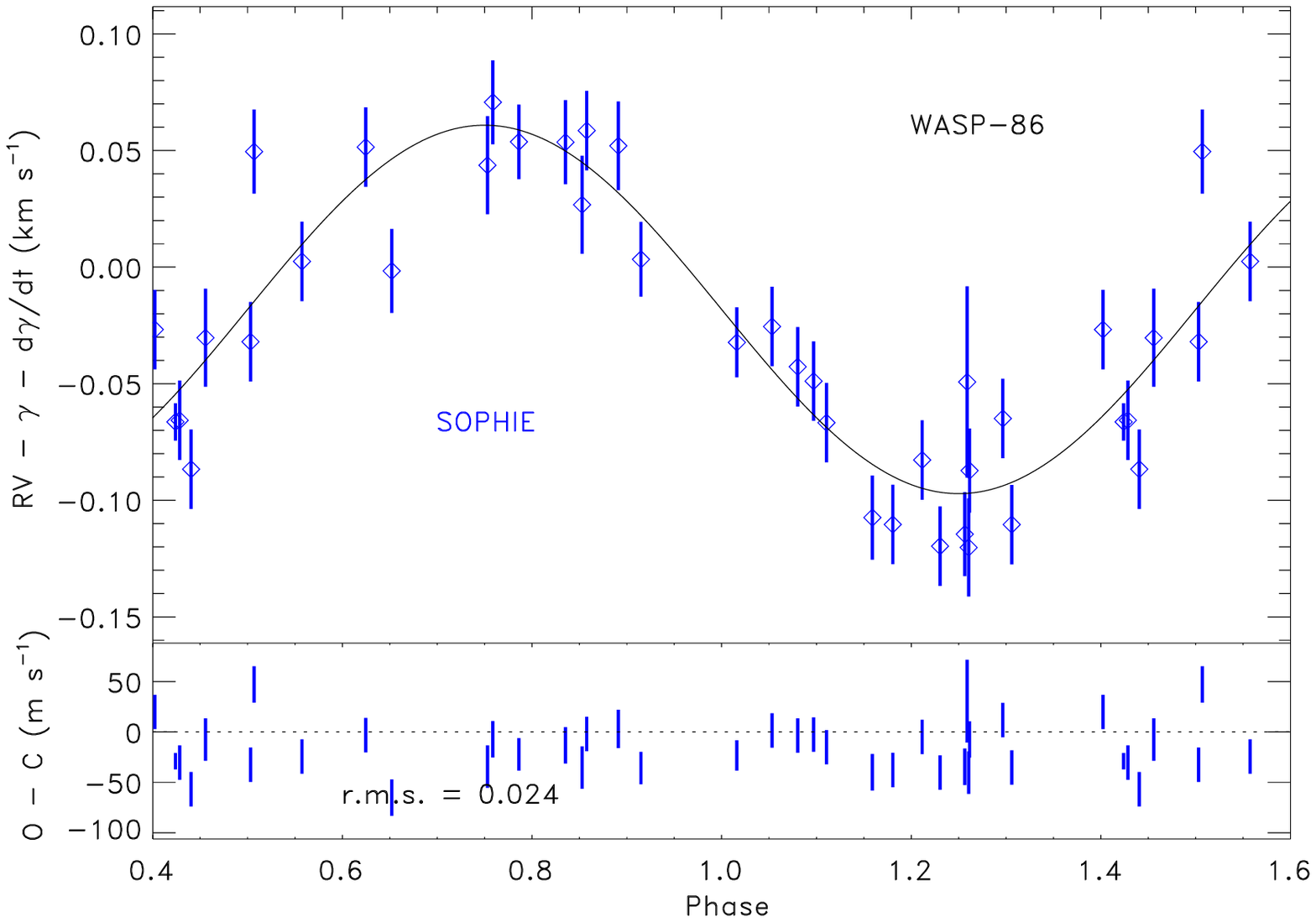}
   \includegraphics[width=0.49\textwidth]{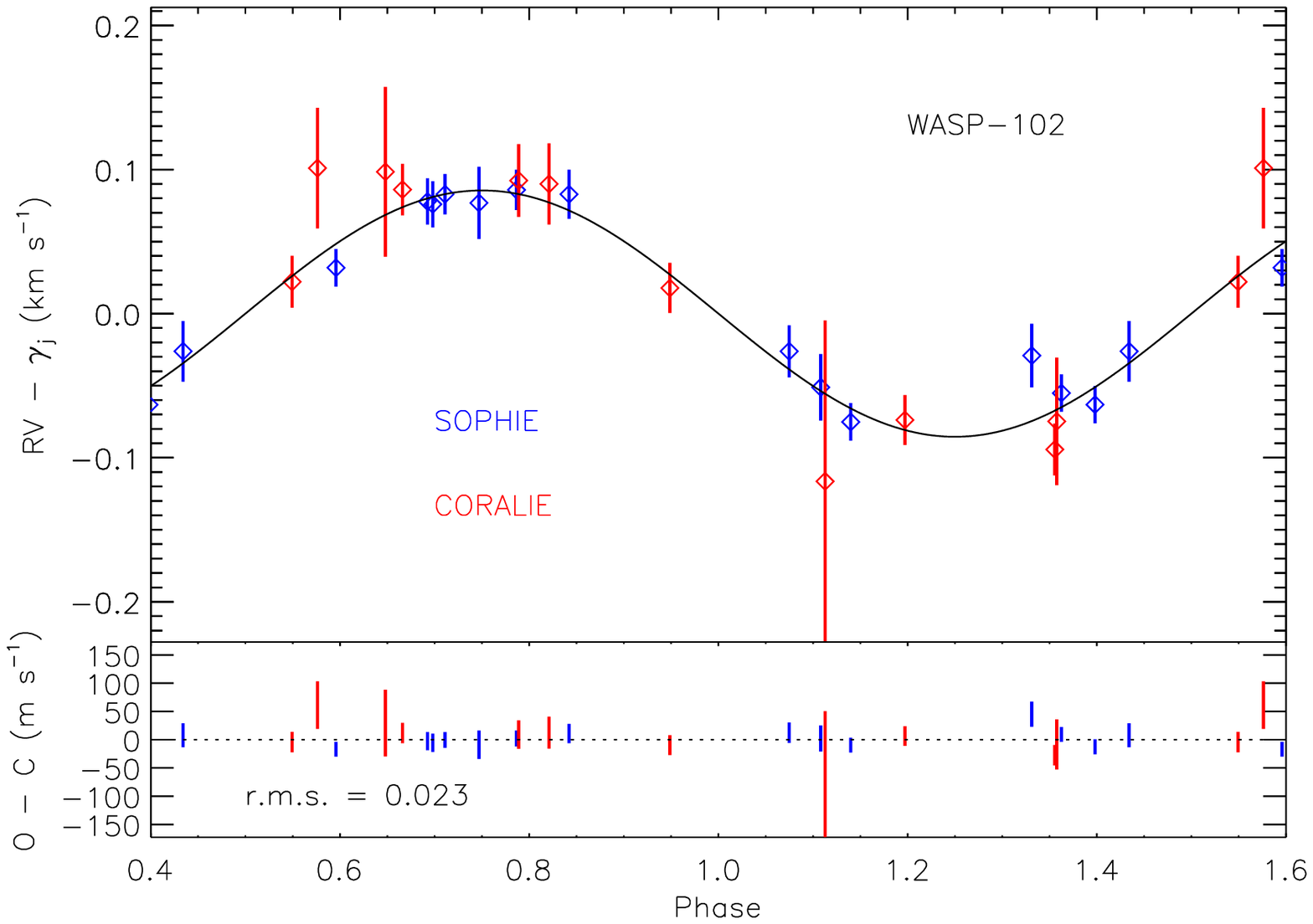}
   \caption{{\emph Upper panels}: Phase folded relative radial velocity
     measurements of WASP-86 (left) and WASP-102 (right) obtained with
     the SOPHIE (blue) and CORALIE (red) spectrographs. The systemic
     velocity and the long-term trend ($d\gamma/dt$) have been
     subtracted from the RVs.  Superimposed is the best-fit model RV
     curve with parameters from Table~\ref{planets_params}. {\emph Lower
       panels}: Residuals from the radial velocity fit plotted against
     orbital phase; the dotted line in the lower panels marks zero.
     The residuals are in units of \ms. The r.m.s.\ are in units of
     \kms.}
    \label{figrvs}%
   \end{figure*}

\begin{figure*}
   \centering
   \includegraphics[width=0.49\textwidth]{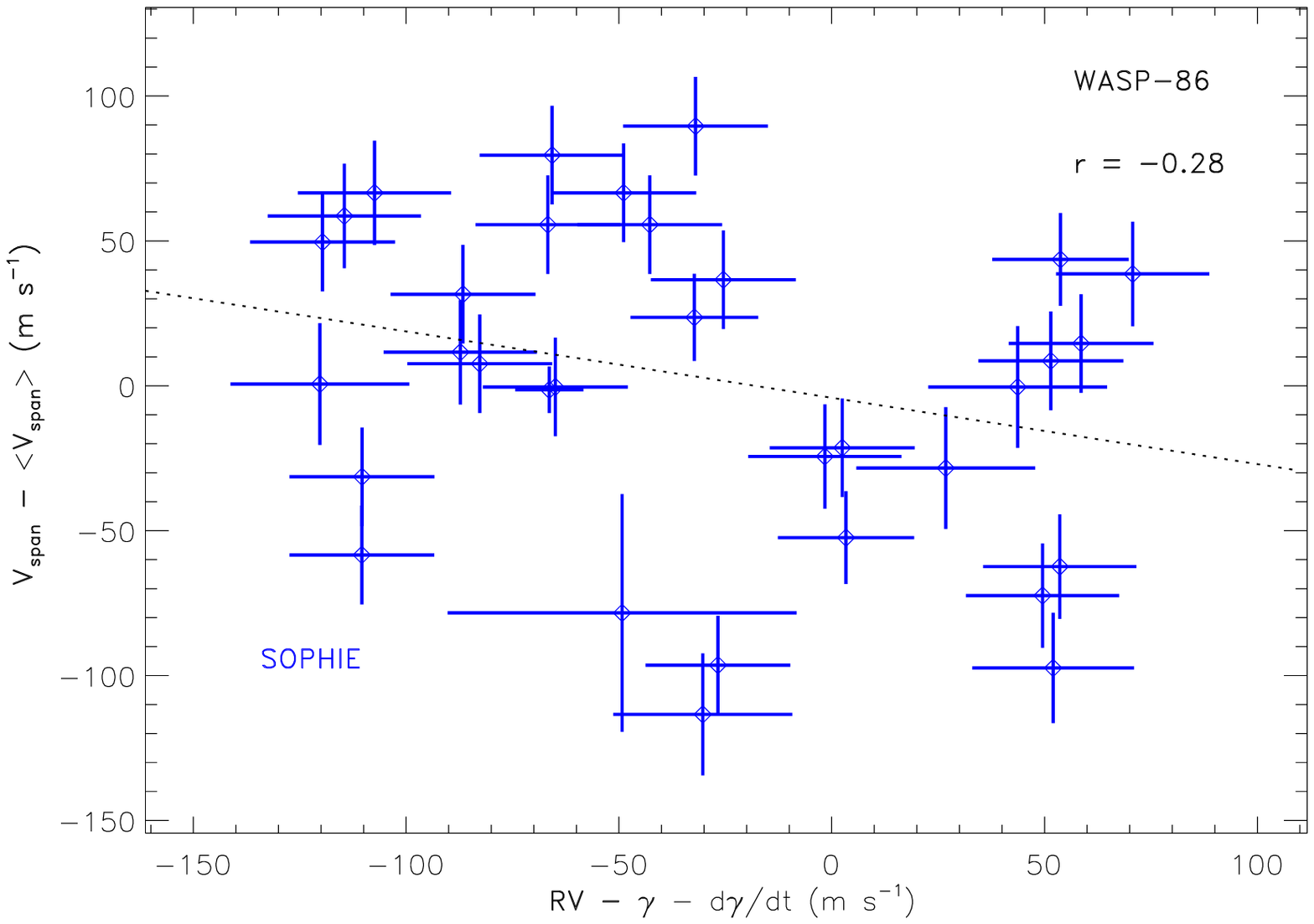}
   \includegraphics[width=0.49\textwidth]{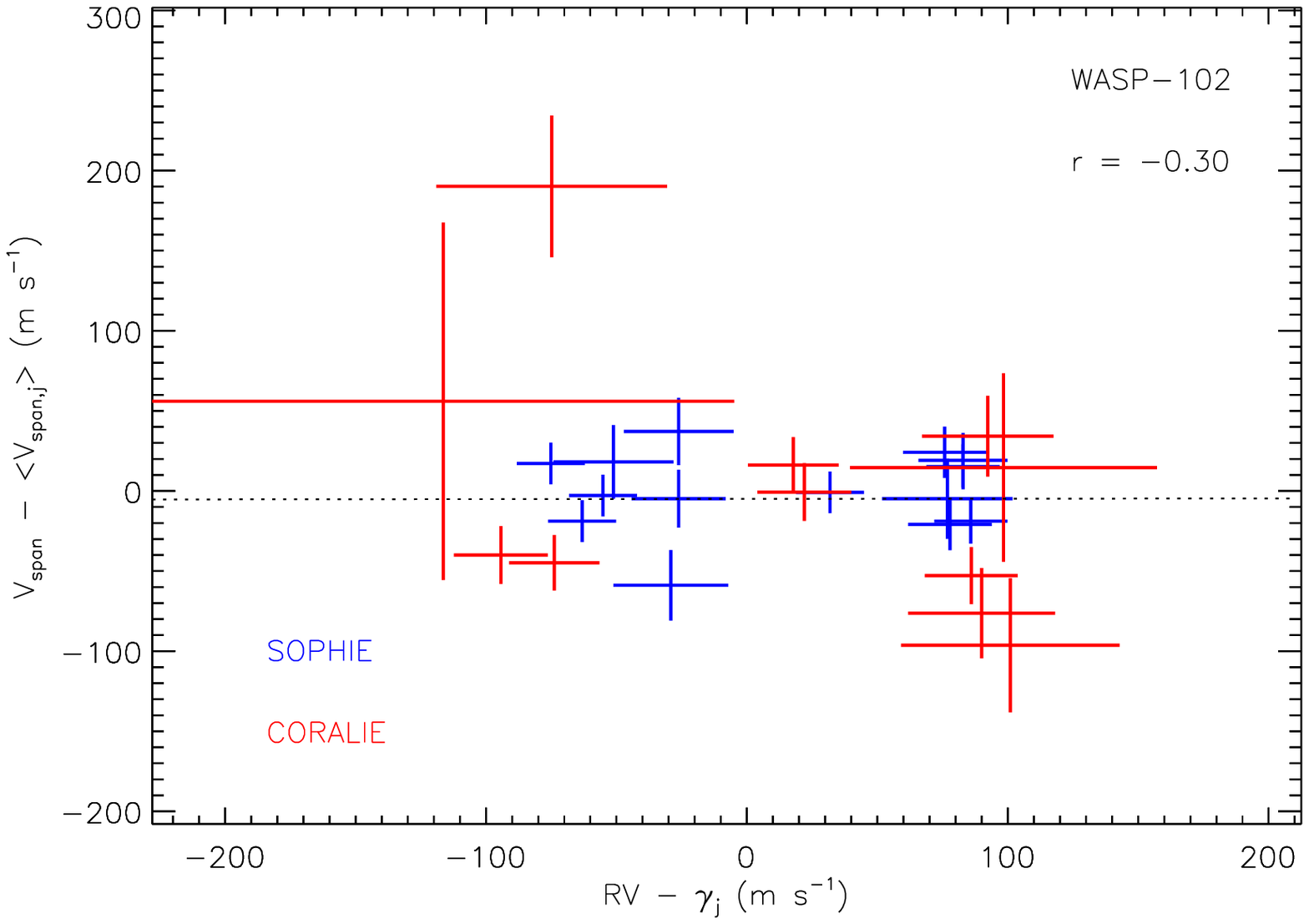}
   \caption{We plot the bisector span measurements of WASP-86 (left)
     and WASP-102 (right) as a function of relative radial
     velocity. From the bisector span value we subtract the mean
     bisector span for each instrument while removing the systemic
     velocity and in the case of WASP-86 the RV long-term trend
     ($d\gamma/dt$).  In the case of WASP-86, $<$V$_{\rm span}>$ =
     -36.9~\ms, and in the case of WASP-102, $<V_{\rm span,SOPHIE}>$ =
     -18.1~\ms\ and $<V_{\rm span,CORALIE}>$ = 11.1~\ms.  The
     horizontal, dotted line denotes the best linear fit to all of the
     data points including the individual uncertainties.  The V$_{\rm
       span}$ measurements of both stars are of the same order of
     magnitude as the errors in the RVs, and show no significant
     variation nor correlation with RVs, as indicated by the Pearson
     product-moment correlation coefficient, $r$.  }
    \label{figvspan}
   \end{figure*}

In Figure \ref{figrvs} we plot the phase folded radial velocity curve
for WASP-86 (left) and WASP-102 (right). Additionally, the RV
residuals from our best fit model are plotted against orbital phase
(Figure \ref{figrvs}: \emph{lower--panel}). The $RMS$ of the residuals
to the best fit Keplerian models are as follows: $RMS = 30$~ms$^{-1}$
for WASP-86, and $RMS = 15$~ms$^{-1}$ for WASP-102, which are
comparable to the errors in the RV measurements.  The systemic
velocity and the long-term trend ($d\gamma/dt$) have been subtracted
from the RVs, which in the case of WASP-86 are $\gamma$ = $-$23.676
$\pm$ 0.015~$\kms$~and~ $d\gamma/d$ = 30.2$\pm$2.7~$\ms$ $y^{-1}$, and
in the case of WASP-102 are \gammaS\ = $-$16.54584 $\pm$ 0.00064~\kms,
\gammaC\ = $-$16.5459 $\pm$ 0.00064 $\kms$ and $d\gamma/dt$ = 0.  In
Figure \ref{figvspan} we plot the bisector span measurements (V$_{\rm
  span}$) for both systems versus radial velocity. The bisector span
measurements of both planet hosts are of the same order of magnitude
as the errors in the RV measurements, and show no significant
variation nor correlation with radial velocity, as indicated by the
Pearson product-moment correlation coefficient, $r$, see Figure
\ref{figvspan}.  This suggests that the radial velocity variations
with semi-amplitudes of $K_1$ = 0.0845 $\pm$0.0052~\kms\ for WASP-86b,
and $K_1$ = 0.0855 $\pm$0.0049~$\kms$ for WASP-102b, are due to
Doppler shifts of the stellar lines induced by a planetary companion
rather than stellar profile variations due to stellar activity or a
blended eclipsing binary.

In the radial velocity signal of WASP-86b there is indication of an
additional body in the system. Our current RV dataset does not allow
us to confirm a third object nor to put constraints on its mass, but
our analysis of the RV signal and bisector allow us to exclude
relatively massive stellar companions.

\begin{table}
  \caption{Radial velocity measurements of WASP-86 obtained with SOPHIE. The columns are: the Baricentric Julian date (BJD-TDB), the stellar RV measurements, the RV uncertainties, the line-bisector span measurements and the residuals to the fit.}
\begin{center}
\begin{tabular}{cccll}
\hline \hline \\
BJD & RV & $\sigma_{\rm RV}$ & V$_{\rm span}$& O -- C\\
$-$2\,400\,000 &(\kms)&(\kms)& (\kms) & (\ms) \\
\hline \\
56063.5320 &    $-$23.800 &    0.018 &  $-$0.028&   $-$7.5 \\
56066.4192 &    $-$23.659 &    0.018 &  $-$0.102&   $-$13.3 \\
56081.6018 &    $-$23.685 &    0.021 &  $-$0.068&   $-$35.3 \\
56084.4756 &    $-$23.778 &    0.008 &  $-$0.041&   $-$29.0 \\
56089.5905 &    $-$23.798 &    0.017 &  $-$0.008&   $-$56.8 \\
56101.3934 &    $-$23.657 &    0.016 &  $+$0.004&   $-$22.2 \\
56103.3771 &    $-$23.821 &    0.017 &  $-$0.071&   $-$37.8 \\
56121.3528 &    $-$23.666 &    0.021 &  $-$0.040&   $-$34.3 \\
56123.3949 &    $-$23.817 &    0.018 &  $+$0.027&   $-$40.0 \\
56125.3998 &    $-$23.707 &    0.017 &  $-$0.061&   $-$24.4 \\
56865.3807 &    $-$23.619 &    0.017 &  $-$0.031&   $-$3.2 \\
56939.2511 &    $-$23.777 &    0.017 &  $-$0.098&   $-$35.2 \\
56940.2622 &    $-$23.617 &    0.018 &  $-$0.112&   $+$47.1 \\
56948.2628 &    $-$23.715 &    0.017 &  $+$0.027&   $-$2.7 \\
56949.2668 &    $-$23.731 &    0.017 &  $-$0.040&   $+$11.7 \\
56950.3063 &    $-$23.698 &    0.017 &  $+$0.050&   $-$32.5 \\
56974.2357 &    $-$23.714 &    0.041 &  $-$0.118&   $+$30.6 \\
56974.2434 &    $-$23.785 &    0.021 &  $-$0.039&   $-$40.4 \\
56975.2256 &    $-$23.695 &    0.021 &  $-$0.153&   $-$7.6 \\
56977.2482 &    $-$23.606 &    0.017 &  $-$0.025&   $-$2.0 \\
56978.2316 &    $-$23.690 &    0.017 &  $-$0.003&   $+$1.4 \\
56979.2545 &    $-$23.779 &    0.018 &  $+$0.019&   $-$34.6 \\
56981.2452 &    $-$23.666 &    0.018 &  $-$0.064&   $-$65.1 \\
57107.5714 &    $-$23.587 &    0.018 &  $-$0.001&   $-$7.2 \\
57133.5154 &    $-$23.653 &    0.016 &  $-$0.092&   $-$35.9 \\
57134.4998 &    $-$23.723 &    0.017 &  $+$0.016&   $-$15.1 \\
57154.4733 &    $-$23.698 &    0.017 &  $+$0.016&   $-$3.6 \\
57158.5529 &    $-$23.603 &    0.019 &  $-$0.137&   $+$3.1 \\
57190.4498 &    $-$23.773 &    0.017 &  $+$0.010&   $-$40.2 \\
57191.4466 &    $-$23.719 &    0.017 &  $+$0.040&   $-$30.4 \\
57210.4813 &    $-$23.735 &    0.017 &  $-$0.032&   $-$5.0 \\
57211.4407 &    $-$23.679 &    0.017 &  $-$0.136&   $+$19.8 \\
57335.2876 &    $-$23.678 &    0.015 &  $-$0.016&   $-$23.3 \\
\bottomrule                                      
\end{tabular}                                   
\label{WASP-86_rvtable}  
\end{center}      
\end{table}

\begin{table}
\caption{Radial velocity measurements of WASP-102. The columns are: 
the Baricentric Julian date (BJD-TDB), the stellar RV measurements, the RV uncertainties, 
the line-bisector span measurements, the residuals to the fit and the instrument 
with which the observations were acquired (S = SOPHIE; C = CORALIE)}
\begin{center}
\begin{tabular}{lcclll}
\hline 
\hline \\
BJD & RV & $\sigma_{\rm RV}$ & V$_{\rm span}$& O -- C & Inst.\\
$-$2\,400\,000 &(\kms)&(\kms)& (\kms)& (\ms) &  \\
\hline \\
    56186.4885 &    $-$16.609 &      0.013 &      $-$0.037 &      $-$12.4 & S \\
    56187.5411 &    $-$16.460 &      0.014 &      $-$0.037 &      $+$2.3 & S \\
    56188.4985 &    $-$16.621 &      0.013 &      $-$0.001 &      $-$9.7 & S \\
    56192.4438 &    $-$16.514 &      0.013 &      $-$0.019&       $-$16.9 & S \\
    56195.4307 &    $-$16.470 &      0.016 &      $+$0.006&      $-$5.5 & S \\
    56213.4903 &    $-$16.601 &      0.013 &      $-$0.021&       $+$9.5 & S \\
    56214.4346 &    $-$16.463 &      0.014 &      $+$0.003&      $-$0.4 & S \\
    56216.3941 &    $-$16.572 &      0.021 &      $+$0.019&      $+$7.9 & S \\
    56270.3117 &    $-$16.575 &      0.022 &      $-$0.077&      $+$45.1 & S \\
    56271.2903 &    $-$16.468 &      0.016 &      $-$0.039&      $-$2.4 & S \\
    56272.3265 &    $-$16.572 &      0.018 &      $-$0.023&       $+$12.2 & S \\
    56285.2449 &    $-$16.463 &      0.017 &      $+$0.001&      $+$10.9 & S \\
    56301.2458 &    $-$16.469 &      0.025 &      $-$0.023&       $-$9.0 & S \\
    56302.2248 &    $-$16.597 &      0.023 &      $-$0.000&      $+$2.2 & S \\
    56536.7314 &    $-$16.382 &      0.059 &     $+$0.107&      $+$29.5 & C \\
    56538.6544 &    $-$16.555 &      0.044 &     $+$0.012&       $-$8.5 & C \\
    56540.7007 &    $-$16.597 &      0.112 &     $-$0.091 &      $-$61.1 & C \\
    56543.6393 &    $-$16.554 &      0.017 &     $+$0.036&       $+$6.6 & C \\
    56544.6662 &    $-$16.379 &      0.042 &     $-$0.104 &      $+$61.3 & C \\
    56545.6756 &    $-$16.463 &      0.017 &     $+$0.028&       $-$9.7 & C \\
    56547.6195 &    $-$16.394 &      0.018 &     $+$0.040&       $+$11.9 & C \\
    56550.6624 &    $-$16.388 &      0.025 &     $+$0.058&       $+$9.1 & C \\
    56595.5549 &    $-$16.575 &      0.018 &     $+$0.023&       $-$27.3 & C \\
    56599.5259 &    $-$16.390 &      0.028 &     $+$0.041&       $+$12.5 & C \\
    56888.7397 &    $-$16.458 &      0.018 &     $-$0.029&       $-$4.3 & C \\
\bottomrule
\end{tabular}
\label{WASP-102_rvtable}
\end{center}
\end{table}

\section{Physical Properties}\label{properties}

We performed our routine analysis on the complete set of spectroscopic
and photometric data for both systems, from which we derive stellar
and planetary physical properties.

\subsection{Spectroscopically-determined stellar properties}\label{star}

The stellar spectroscopic properties for WASP-86 and WASP-102 were
obtained using the co-added spectra from SOPHIE and the methods given
in \citet{Doyle2013}. The excitation balance of the Fe~{\sc i} lines
was used to determine the effective temperature (\teff). The surface
gravity ($\logg$) was determined from the ionisation balance of
Fe~{\sc i} and Fe~{\sc ii}. The Ca~{\sc i} line at 6439{\AA} and the
Na~{\sc i} D lines were also used as $\logg$ diagnostics. Values of
microturbulence (\mictrb) were obtained by requiring a null-dependence
on abundance with equivalent width. The elemental abundances were
determined from equivalent width measurements of several unblended
lines. The quoted error estimates include that given by the
uncertainties in $\teff$ and $\logg$, as well as the scatter due to
measurement and atomic data uncertainties. The projected stellar
rotation velocity ($\vsini$) was determined by fitting the profiles of
several unblended Fe~{\sc i} lines. Macroturbulence was obtained from
the calibration by \cite{Bruntt2010b}. The overall precision of the
parameters was limited by the quite modest signal-to-noise ratios of
the spectra $\sim$80:1 for WASP-86 and $\sim$60:1 for WASP-102.

For WASP-86, the rotation rate ($P_{\rm rot} = 5.7 \pm 0.8$~d) implied
by the {\vsini} gives a gyrochronological age of
$0.8^{+0.8}_{-0.4}$~Gyr using the \citet{Barnes2007} relation.  The
effective temperature of this star is close to the lithium-gap,
\citep{Bohm-Vitense2004}, thus the lack of any detectable lithium does
not provide a usable age constraint.  WASP-86's rotation period,
albeit assuming that the stellar rotation axis is along the plane of
the sky, is remarkably close (within 1--$\sigma$ from \vsini) to the
orbital period of WASP-86b P = 5.03~days.  Using the long baseline of
WASP data (covering $\sim$6 yrs) we have investigated the presence of
photometric modulation in the WASP light curve due to stellar
variability. However, we do not detect any variation down to 0.5~mmag
(see also \S \ref{swobs} and Fig.~\ref{periodo}).  A more detailed
analysis of the spectrum revealed no sign of stellar activity with a
S/N of about 20:1 in the core of the Ca II H\&K lines.  The stellar
mass and radius were estimated using the calibration of
\citet{Torres2010}.

In the case of WASP-102, the rotation rate ($P_{\rm rot} = 8.1 \pm
1.5$~d) implied by the {\vsini} gives a gyrochronological age of
$0.6^{+0.5}_{-0.3}$~Gyr using the \citet{Barnes2007} relation. An
estimated lithium age of $\sim$0.5--2 Gyr estimated using
\citet{Sestito2005} is consistent. As in the case of WASP-86, there is
no indication of stellar activity for WASP-102.

\begin{table}
\caption{Stellar Parameters from spectral analysis.}
\begin{tabular}{cccc} \hline \hline \\
Parameter           & WASP-86       & WASP-102      \\ \hline \\
\teff (K)           & 6330$\pm$110  & 5940$\pm$140  \\
\logg               & 4.28$\pm$0.10 & 4.49$\pm$0.11 \\
{[Fe/H]}            &+0.23$\pm$0.14 &+0.09$\pm$0.19 \\
\mictrb (\kms)      & 1.1$\pm$0.2   & 1.0$\pm$0.2   \\
\mactrb (\kms)      & 4.3$\pm$0.3   & 3.0$\pm$0.3   \\
\vsini (\kms)       & 12.1$\pm$0.8  & 6.2$\pm$0.7   \\
$\log A$(Li)        & $<$1.08       & 2.44$\pm$0.12 \\
Mass (\msun)   & 1.32$\pm$0.11 & 1.10$\pm$0.10 \\
Radius (\rsun) & 1.36$\pm$0.18 & 1.00$\pm$0.14 \\
Sp. Type            & F7            & G0            \\
Age                 & $0.8^{+0.8}_{-0.4}$~Gy&   $0.6^{+0.5}_{-0.3}$~Gyr     \\
\hline 
\\
\end{tabular}
\label{wasp-params}
\newline {\bf Notes:} 
Macroturbulence (\mactrb) was obtained from the
calibration by \citet{Bruntt2010b}. 
The mass and radius were obtained using the \citet{Torres2010} calibration. 
The spectral type is estimated from \teff\ 
using the table in  \citet{Gray2008}.
The iron abundance is relative to
the solar value obtained by \citet{Asplund2009}. 
\end{table}

\subsection{Stellar masses and ages}\label{mass-age}

For both systems we used stellar theoretical evolutionary models in
order to estimate stellar ages and masses.  We used the stellar
densities \rhostar, measured directly from our Markov-Chain Monte
Carlo (MCMC) analysis (\S \ref{mcmc}, and see also
\citealt{Seager2003}), together with the stellar temperatures and
metallicity values derived from spectroscopy, to perform an
interpolation over three different stellar evolutionary models. For
details of the interpolation method see \citet{Brown2014}.

The stellar density, $\rhostar$, is directly determined from transit
light curves and as such is independent of the effective temperature
determined from the spectrum \citep{Hebb2009}, as well as of
theoretical stellar models (if $\mpl \ll \mstar$ is assumed). The
following stellar models were used: a) the Padova stellar models
(\citealt{Marigo2008}, and \citealt{Girardi2010}), b) the Yonsei-Yale
(YY) models \citep{Demarque2004}, and c) models from the Dartmouth
Stellar Evolution Program (DSEP)
(\citealt{Chaboyer2001},\citealt{Bjork2006}, \citealt{Dotter2008}).

\begin{figure*}
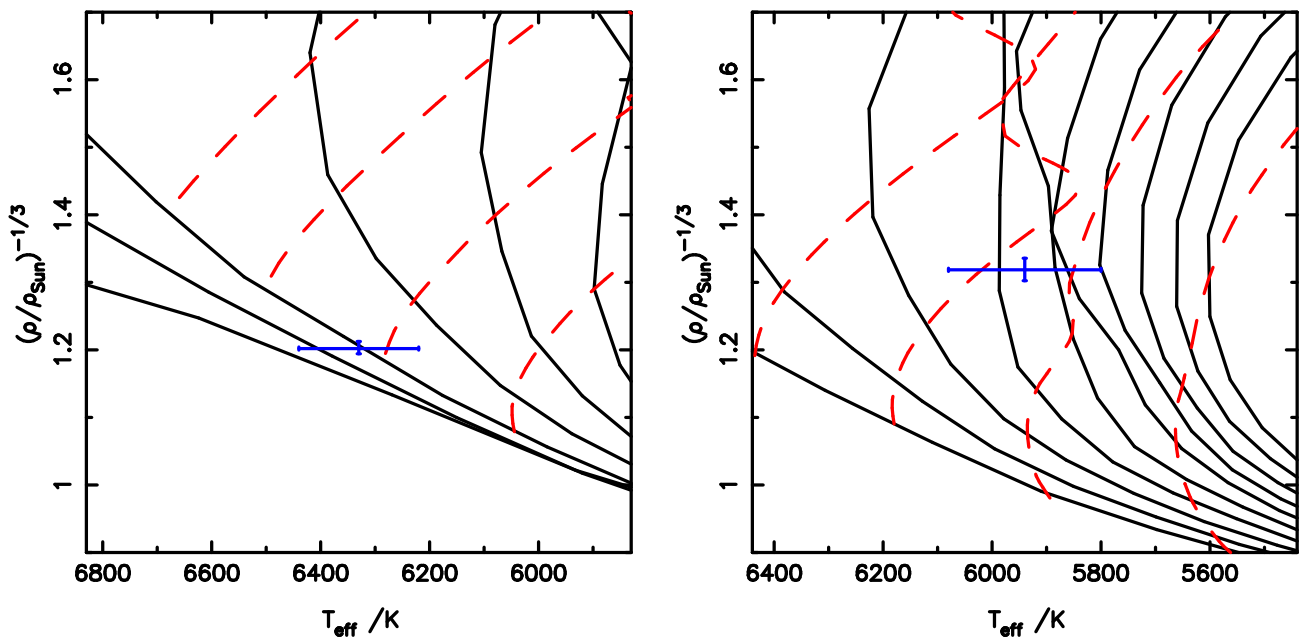

  \centering
   \includegraphics[width=0.45\textwidth]{W86_YYFeH+023.eps}
\hspace{0.3cm}
\includegraphics[width=0.45\textwidth]{Wasp102_yonseiy.eps}
\caption{Isochrone tracks from \citet{Demarque2004} for WASP-86 and
  WASP-102 using the metallicity \feh $=+0.23$ dex and \feh $=+0.09$
  dex respectively from our spectral analysis and the best-fit stellar
  density \rhostar.  \emph{Left-panel}: WASP-86, from left to right
  the solid lines are for isochrones of: 0.1, 0.6, 1, 2, 3, and
  4~Gyr. From left to right, dashed lines are for mass tracks of: 1.4,
  1,3, 1.2, and 1.1~\msun. \emph{Right-panel}: WASP-102, from left to
  right the solid lines are for isochrones of: 1, 2, 3, 4, 5, 6, 7, 8,
  9, and 10 Gyr. From left to right, red--dashed lines are for mass
  tracks of: 1.3, 1.2. 1.1. and 1.0~\msun. }
    \label{iso1}
   \end{figure*}

   In Figure \ref{iso1} we plot the inverse cube root of the stellar
   density \rhostar$^{-1/3} =$~\rstar/\mstar$^{1/3}$ (solar units)
   against effective temperature, \teff, for the selected model mass
   tracks and isochrones for the two planet host stars
   respectively. For WASP-86 the stellar properties derived from the
   three sets of stellar evolution models (Table~\ref{tracks}) agree
   with each other, and with those derived from the \citet{Torres2010}
   calibration, within their 1--$\sigma$ uncertainties.

   This is not the case for WASP-102 for which the stellar ages
   derived from theoretical models are quite different from the age
   obtained via gyrochronology see Table \ref{tracks}.  In \S
   \ref{star} from \vsini~and using \citet{Barnes2007} relation we
   obtained for WASP-102 an estimated stellar rotation period of about
   8~days. For a G0 star this translate to an age of
   $\sim$0.6~Gyr. Moreover, the age estimate from the lithium
   abundance also suggests a young age (0.5--2~Gyrs).  These values
   are in disagreement with the three evolutionary model estimates
   obtained above. Their 1--$\sigma$ lower limits seem to indicate an
   age of 2.6~Gyrs or older. However, we note that these values are
   fairly unconstrained and have large errors.  Additionally,
   different sets of theoretical models might not perfectly agree with
   each other \citep{Southworth2010}, and moreover at younger ages
   isochrones are closely packed and a small change in \teff\ or
   \rhostar\ can have a significant effect on the derived stellar
   age. This is visible in Figure \ref{iso1} where at young stellar
   ages isochrones are practically indistinguishable.

   In principle, isochrone fitting is applicable to stars across the
   spectral range, but it can be difficult to determine ages for stars
   with spectral type later than mid-to-late G owing to the fact that
   they evolve very slowly, having nuclear burning time-scales that
   are longer than the age of the Galactic disc.  Given our error on
   the \teff, \feh\ and \rhostar\ this could possibly explain the age
   discrepancy of WASP-102.
 
   Even with precise stellar densities problem can arise.  Grids of
   stellar models use a broad sampling in mass, age and metallicity
   that can produce poor sampling of the observed parameter
   spacing. In such a case the difference in stellar density between
   adjacent grid points can be much larger than the uncertainty on the
   observed value. This can produce systematic interpolation errors
   and make reliable estimates of the uncertainties on the mass and
   age difficult. Moreover, some combinations of mass, age and
   composition could be missed because are not sampled by the stellar
   model grid or fall just outside the 1--$\sigma$ error bars,
   particularly when fitting by-eye.

\begin{figure}
   \centering
   \includegraphics[width=0.49\textwidth]{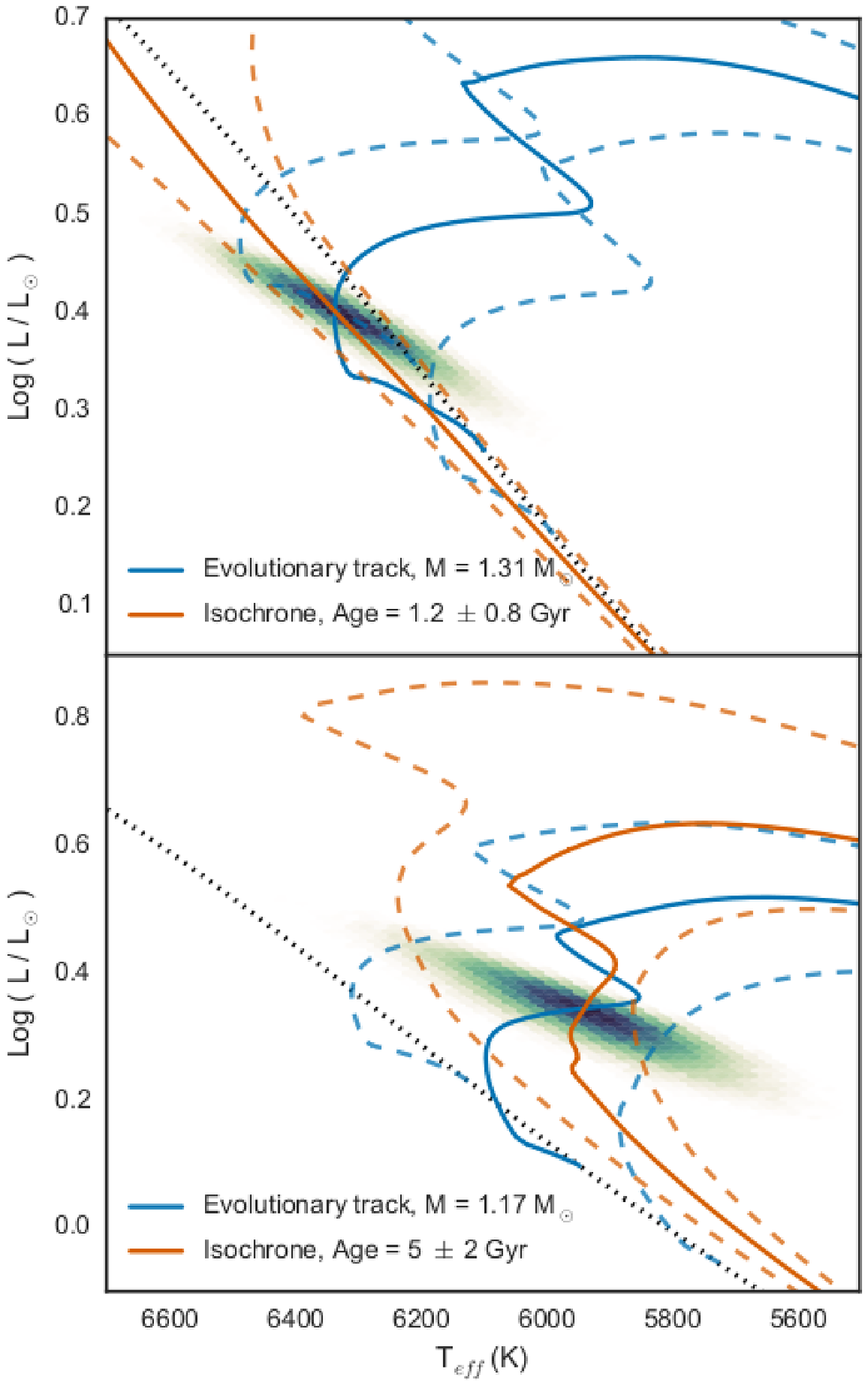}
   \caption{BAGEMASS stellar mass and age analysis of WASP-86
     (\emph{upper panel}) and WASP-102 (\emph{lower panel}). The
     dotted black line is the ZAMS. The solid blue line is the mass
     evolutionary track, and the blue dashed tracks on either side are
     for the 1--$\sigma$ error of the mass. The solid orange line is
     the stellar age isochrone and the orange dashed lines represents
     the 1--$\sigma$ error. The density of MCMC samples is shown in
     the colour scale of the posterior distribution plotted.  }
    \label{iso2}
   \end{figure}

   To avoid any such problem we have used the approach of
   \citet{Maxted2015b} to evaluate the masses and ages for both
   systems. \citet{Maxted2015b} developed a Markov-Chain Monte Carlo
   (MCMC) method (BAGEMASS), a Bayesian method that calculates the
   posterior probability distribution for the mass and age of a star
   from its observed mean density and other observable quantities
   (e.g. \teff, \feh\ and luminosity) using a grid of stellar models
   that densely samples the relevant parameter space.  In Table
   \ref{tracks} and Figure \ref{iso2} we show the masses and ages
   obtained with this method. The results from BAGEMASS confirm a
   young age for WASP-86 of about 1.2$\pm$0.8~Gyr, and also find a
   mass estimate for WASP-86 (1.31$\pm$0.06 \msun) in agreement with
   our previous estimate. For WASP-102 instead we obtain an older age
   (5$\pm$2~Gyr) compared to gyrochonology in agreement with our
   previous estimate from stellar models, while the estimated stellar
   mass of 1.17$\pm$0.09 \msun~is in agreement with that derived from
   stellar models above and that from empirical calibrations
   (e.g. \citealt{Torres2010}).

\begin{table*}
\caption{Stellar masses and ages for WASP-86 and WASP-102.} 
\begin{center}
\begin{tabular}{lllll} 
\hline\hline \\ 
 & \multicolumn{2}{c}{WASP-86} & \multicolumn{2}{c}{WASP-102} \\
\midrule  
\medskip
 & \mstar~(\msun) & Age (Gyr) & \mstar~(\msun) & Age (Gyr) \\
 \cline{2-5}\\ 
Padova$^a$ & 1.29$^{+0.04}_{-0.09}$&0.52$^{+1.54}_{-0.46}$&1.12$^{+0.10}_{-0.15}$&5.35$^{+5.35}_{-2.16}$ \smallskip\\
YY$^b$     & 1.31$^{+0.08}_{-0.08}$&1.05$^{+1.16}_{-0.84}$&1.18$^{+0.04}_{-0.19}$&4.45$^{+4.85}_{-1.76}$ \smallskip\\
DSEP$^c$   & 1.21$^{+0.07}_{-0.05}$&2.72$^{+0.80}_{-1.38}$&1.08$^{+0.16}_{-0.15}$&6.49$^{+4.98}_{-3.05}$ \medskip \\
\midrule 
BAGEMASS$^d$ & 1.31$\pm$0.06&1.2$\pm$0.8&1.17$\pm$0.09&5$\pm$2 \smallskip\\
\midrule
\end{tabular}
\label{tracks}
\end{center} {\small $^a$~\citet{Marigo2008,Girardi2010};
  $^b$~\citet{Demarque2004};
  $^c$~\citet{Chaboyer2001,Bjork2006,Dotter2008} };
  $^d$~\citet{Maxted2015a}
\end{table*} 
Finally, considering the possible range of ages within the 1--$\sigma$ uncertainties
we adopted an age of 1.2$\pm0.8$~Gyr and 5$\pm2$~Gyr for WASP-86 and
WASP-102, respectively.  In Figures \ref{iso2}, we show for each
planet host star a plot with the chosen set of stellar models from
\citet{Maxted2015a}, while we give a comprehensive list of all models
results in Table~\ref{tracks}.

\begin{figure*}
   \centering
   \includegraphics[width=0.7\textwidth]{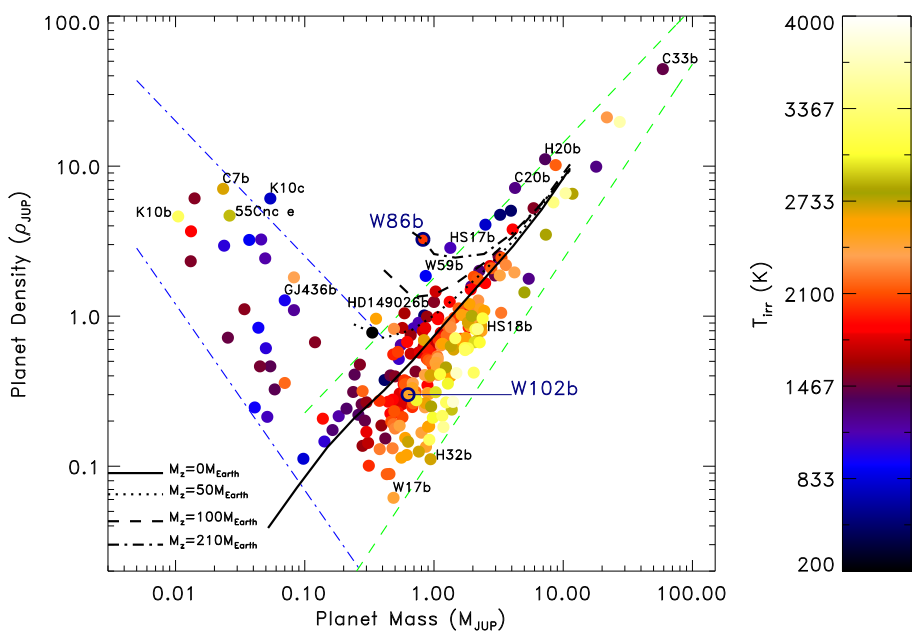}
   \caption{Planetary density versus mass and irradiation temperature
     (colours) for all planets with precisely derived masses and radii
     (better than 20\%). We also plot boundaries in this plane
     (green--dashed and blue--dot-dashed lines) as derived in
     \citet{Bakos2015} but we estimate different values which are
     described in \S\ref{discussion}. In black, we show theoretical
     models of planetary interiors \citep{Fortney2007,Baraffe2008} for
     different core masses: 0\me~solid line, 50\me~dotted line,
     100\me~dashed line, and 210\me~dot-dashed lines. The latter is
     obtained by extrapolation from the models above. The loci of
     WASP-86b and WASP-102b are indicated in navy blue. WASP-86b is a
     clear outlier and is the most dense planet among planets with
     masses between that of Neptune and 2\mj, making it quite
     exceptional. Other planets are indicated by their name in full or
     with the following criteria: K for {\it Kepler}, H for HAT, HS
     for HAT-South, C for CoRoT and W for WASP.  }
    \label{mass-density}
   \end{figure*}

\subsection{Planetary physical properties}\label{mcmc}

The planetary properties were determined using our thoroughly tested
Markov-Chain Monte Carlo (MCMC) analysis, which we performed including
all available WASP and follow-up photometry, together with SOPHIE and
CORALIE radial velocity measurements (as appropriate for each
system). A detailed description of the method is given in
\citet{Cameron2007} and \citet{Pollacco2008}.

In our iterative fitting we used the following jump parameters: the
epoch of mid transit $T_{0}$, the orbital period $P$, the fractional
change of flux proportional to the ratio of stellar to planet surface
areas $\Delta F = R_{\rm pl}^2/R_{\star}^2$, the transit duration
$T_{14}$, the impact parameter $b$, the radial velocity semi-amplitude
$K_{\rm 1}$, the stellar effective temperature \teff\ and metallicity
\feh, the Lagrangian elements \secos\ and \sesin~(where $e$ is the
eccentricity and $\omega$ the longitude of periastron), and the
systemic offset velocity $\gamma$.  For WASP-102 we fitted the two
systemic velocities $\gammaC$ and $\gammaS$ separately to allow for
instrumental offsets between the two data sets. In the case of WASP-86
we also fitted for a long term trend in the radial velocities
($d\gamma/dt$).  The sum of the $\chi^2$ for all input data curves
with respect to the models was used as the goodness-of-fit statistic.
For each planetary system four different sets of solutions were
considered: with or without the main-sequence mass-radius constraint
(MS constraint) and in the case of circular orbits or orbits with
floating eccentricity.

An initial MCMC solution with a linear trend in the systemic velocity
as a free parameter was explored for the two planetary systems. In the
case of WASP-86 a linear trend to the RV data was found to be
significant. When exploring solutions with non-circular orbits the
Lucy \& Sweeney F-test was performed \citet{Lucy1971}, and in both
cases this returned a probability of 100\% that the improvement in the
fit produced by the best-fitting eccentricity could have arisen by
chance if the orbit were in fact circular. For the treatment of the
stellar limb-darkening, the models of \citet{Claret2000, Claret2004}
were used in the $r$-band, for WASP, SPM, NITES, LT and Euler
photometry, and in the $z$-band for TRAPPIST and FTN photometry. At
each MCMC chain step we look-up the limb-darkening coefficients from
these tabulations using the value of \teff\ for that step assuming
\logg as derived in Table \ref{wasp-params}.

We calculate the mass $\mstar$, radius $\rstar$, density $\rhostar$,
and surface gravity $\logg$ of the host stars as well as \mpl~, \rpl~,
\rhopl~ and $\logg_{\rm{pl}}$~ for the planets and their the
equilibrium temperatures assuming a black-body ($T_{\rm pl,A=0}$) and
efficient energy redistributed from the planet's day-side to its
night-side. We also calculate the transit ingress/egress times $T_{\rm
  12}$/$T_{\rm 34}$, and the orbital semi-major axis $a$. All
calculated values and their 1--$\sigma$ uncertainties are presented in
Table~\ref{planets_params}. The corresponding best-fitting transit
light curves are shown in Figures~\ref{W86_lcs} and
\ref{W102_lcs}. The best-fitting RV curves are presented in
Figure~\ref{figrvs}.

For both planets, circular orbits and no main-sequence constraint on
the stellar mass and radius were selected. We find that imposing the
MS constraint has little effect on the MCMC global solution in the
case of WASP-86, while in the case of WASP-102 imposing the MS
constraint was increasing the posterior probability values for the
stellar temperature and metallicity beyond their 1--$\sigma$ spectral
uncertainties as derived in our analysis in \S \ref{star}.

\begin{figure*}[ht]
   \centering
   \includegraphics[width=0.7\textwidth]{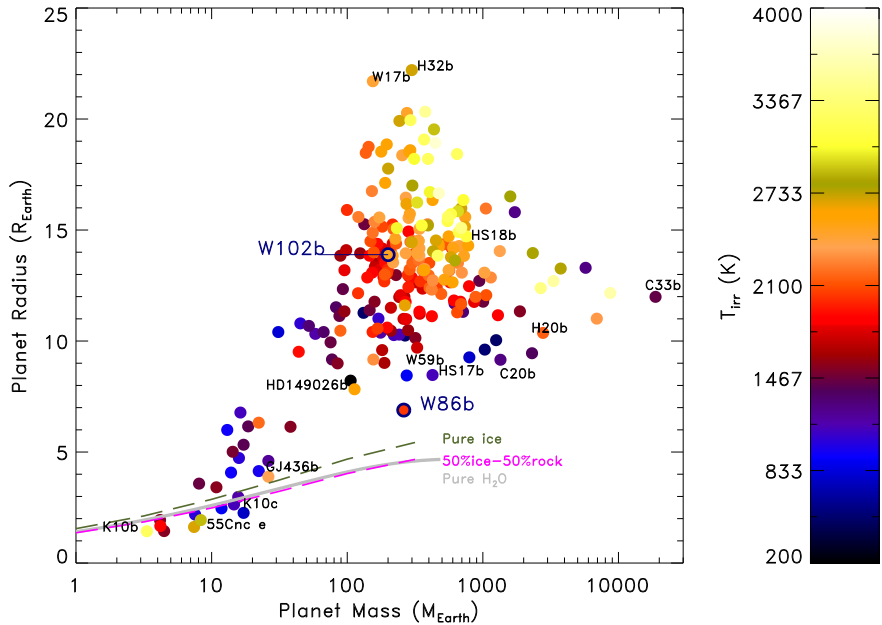}
   \caption{Planetary radius versus mass and irradiation temperature
     (colours) as in Figure \ref{mass-density}. We also plot models of
     planet interior for three different compositions: pure ice (olive
     dashed line), 50\% water -- 50\% rock (magenta dashed line)
     models from \citet{Fortney2007}, and pure water (grey solid line)
     models are from \citealt{Zeng2013}. The loci of WASP-86b and
     WASP-102b are indicated in blue circle and label.  We note that
     WASP-86b is remarkably close to the right--bottom corner where
     planets are forbidden simply because they would be too massive
     for their host star to form \citep{Mordasini2012}. .  Other
     planets are indicated by their name in full or with the following
     criteria: K for {\it Kepler}, H for HAT, HS for HAT-South, C for
     CoRoT and W for WASP. All planets had parameters derived to
     better than 20\%.  }
    \label{mass-radius}
   \end{figure*}

\begin{table*}[t]
\caption{System Parameters of WASP-86 and WASP-102}
\label{planets_params}
\begin{center}
\begin{tabular}{lccc}
\hline\hline \\
&WASP-86&WASP-102& \\
\hline
\\
$P$                     &   5.031555$^{+0.000002}_{-0.000002}$ \smallskip   &2.709813$^{+0.000005}_{-0.000004}$ \smallskip&d\\
$T_{0}~^{\dagger}$      &   7103.7527$^{+0.0004}_{-0.0006}$ \smallskip      &6510.5872$\pm$0.0002                        &d\\
$T_{\rm 14}~^{\ddagger}$&   0.1674$^{+0.0015}_{-0.0010}$ \smallskip         &0.1459$\pm$0.0005                           &d\\
$\Delta F=\rpl^{2}/\rstar^{2}$&    0.0025307$\pm$0.00008                  &0.00946$\pm$0.00010                           & \\
$b$                     &          0.027$^{+0.036}_{-0.020}$ \smallskip   &0.03$^{+0.05}_{-0.02}$ \smallskip             &\rstar\\
$i$                     &         89.85$^{+0.11}_{-0.20}$ \smallskip      &89.73$^{+0.19}_{-0.145}$ \smallskip           &$^\circ$\\
$K_{\rm 1}$             &          0.084$\pm$0.005                        &0.082$\pm$0.006                               &\kms\\
$\gamma$                &        -23.658$^{+0.006}_{-0.005}$ \smallskip   &-16.5453$\pm$0.0007                           &\kms\\
$\gamma_2$              &           ...                                   &-16.4803$\pm$0.0002                           &\kms\\
$d\gamma/dt$            &          0.0000823$^{+0.0000089}_{-0.0000082}$ \smallskip&0. (fixed)                           &\kms$d^{-1}$\\
$e$                     &          0. (fixed)                             &0. (fixed)                                    & \\
\mstar                  &          1.239$\pm$0.028                        &1.167$\pm$0.035                               &\msun\\
\rstar                  &          1.291$^{+0.014}_{-0.013}$ \smallskip   &1.331$^{+0.013}_{-0.012}$ \smallskip          &\rsun\\
$\log g_{\star}$        &          4.309$\pm$0.006                        &4.257$\pm0.006$                               &cgs\\
\rhostar                &          0.57$\pm$0.01                          &0.419$^{+0.017}_{-0.009}$ \smallskip          &\rhosun\\
\mpl                    &          0.821$\pm$0.056                        &0.624$\pm$0.045                               &\mj\\
\rpl                    &          0.632$^{+0.014}_{-0.013}$ \smallskip   &1.259$\pm0.016$                               &\rj\\
$\log g_{\rm pl}$       &          3.67$\pm$0.03                          &2.95$\pm$0.03                                 &cgs\\
\rhopl                  &          3.24$^{+0.31}_{-0.26}$ \smallskip      &0.311$^{+0.024}_{-0.022}$ \smallskip          &\rhoj\\
a                       &          0.0617$\pm$0.0005                      &0.0401$\pm$0.0004                             &AU\\
\teq                    &          1415$\pm$22                            &1705$\pm$32                                   &K\\
\bottomrule
\end{tabular}
\end{center}
{\small $^{\dagger}$ BJD --~2\,450\,000.0}\\
{\small $^{\ddagger}$ $T_{\rm 14}$: transit duration, time between 1$^{\rm st}$ and 4$^{\rm th}$ contact}\\
\end{table*}

\section{Discussion}\label{discussion}
\subsection{Comparative exoplanetology}
WASP-86 is a young F7 star of about only 1~Gyr old.  With a radius of
$\rpl$ = 0.632~$\rj$ and a mass $\mpl$ = 0.821~$\mj$, WASP-86b has a
density of $\rhopl$ =3.24~$\rhoj$, and is thus 
the densest planet with mass in the range 0.05\mj\ (Neptune mass) $<
\mpl <$ 2~\mj.  Only giant planets with masses larger than about
2$\mj$ have larger densities, making WASP-86b exceptional in its mass
range.  Figure \ref{mass-density} shows the planetary density (\rhoj)
versus mass (\mj) and irradiation temperature as derived in
\citet{Heng2012}, for all planets with precisely derived masses and
radii (better than 20\%)\footnote{data from http://exoplanet.eu/}.  We
show the loci of WASP-86b and WASP-102b in navy blue. The black curves
are planet structure models from \citet{Fortney2007} with core masses
of 0 (solid line), 50$\me$ (dotted line), 100$\me$ (dashed line), and
210$\me$ (dot-dashed line).  Figure \ref{mass-density} also shows
upper and lower boundaries in this plane following the functional form
of \citet{Bakos2015}. The values of the power law we obtain are
however different from those derived in \citet{Bakos2015} and are as
follows:

\[ \rm{For}~\mpl < 0.4\mj\,, \, 
\begin{cases}
      \rhopl < 10^{-2.40}\times (\mpl/\mj)^{-1.24}     \\
    \rhopl > 10^{-0.50}\times (\mpl/\mj)^{-0.90} \\
  \end{cases}
\]

\[\rm{For}~\mpl > 0.4 \mj\,, \,
\begin{cases}
      \rhopl < 1.8\times (\mpl/\mj)^{0.90}     \\
    \rhopl > 0.12\times (\mpl/\mj)^{1.30} \\
  \end{cases}
\]
WASP-86b is a clear outlier. Similar objects with lower densities are
HD~149026b \citep{Sato2005}, WASP-59b \citep{Hebrard2013} and HATS-17b
\citep{Brahm2016}.
 
\subsection{Likely disc mass}
With a stellar mass of $\sim$1.3~$\msun$ and an \feh\ = $+0.23$ dex,
WASP-86 is predicted to possess a higher content of heavy elements,
and thus possibly host additional massive planets
\citep{Mordasini2012}. The maximal planetary mass and disc gas-mass
are correlated because gas giant planets (in the core accretion
scenario) accrete most of their mass in a regime where their accretion
rate is proportional to the disc gas-mass
\citep{Guillot2006,Burrows2007,Miller2011}. This correlation implies
that the metallicity of the system determines the maximum mass a
planet can grow up to in a given disc \citep{Mordasini2012}.  The
total amount of heavy elements (M$_Z$) available in the protoplanetary
disc during planet formation is given by $ \rm{M_{z}} =
\rm{M_{Disc}}\, \, Z_{\odot}\, \, 10^{\feh} $
\citep{Mordasini2014,Baraffe2008}, where Z$_\odot$ is 0.015. M$_{\rm
  Disc}$ is the maximum mass of a stable protoplanetary disc which is
M$_{\rm Disc} \lesssim 0.1\rm{\mstar}$ \citep{Ida2004,Mordasini2012}.
In the case of WASP-86 the maximum heavy element content present in
the disc was:
\[\rm{M_{z}} = 0.1\times1.24\msun\times0.015~10^{+0.23} (3.3~10^{5}~\me/\msun)\simeq 10^3\me \]
According to models of planet formation and migration up to
$\sim$~30\% of the heavy elements content of the protoplanetary disc
can be accreted onto planets \citep{Alibert2005a,
  Alibert2006,Mordasini2009,Mordasini2012,Dodson-Robinson2009}. For
WASP-86b the maximum mass of heavy elements available in the disc
amounts to about 330\me.

\subsection{Heavy metal content}
In order to match WASP-86b's high density, an extrapolation from the
theoretical models of \citet{Fortney2007} and \citet{Baraffe2008}
predicts a planet composition of more than 80\% heavy elements (either
confined in a core or mixed in the envelope). If confined in a core,
this value corresponds to a core mass of approximately 210$\me$
\citep{Fortney2007} for a planet with mass $\mpl \sim260~\me$; Figure
\ref{mass-density} shows theoretical models of the planetary interior
from \citet{Fortney2007}.  WASP-86b's heavy element enrichment is
quite close to the above derived 30\% upper limit of accretion
efficiency.  For models such as those of \citet{Baraffe2008} which
consider the case of heavy elements mixed in the envelope (a more
realistic model for gas giant interiors), the maximum difference in
the planet radius is about 10-12\% smaller compared to the radius
estimated by \citet{Fortney2007}. Thus a lower amount of heavy
elements is needed to match the planet radius for the same mass. Using
the models by \citet{Baraffe2008} we estimate a mass fraction of heavy
elements of $\gtrsim$80\% for WASP-86b.  Planets with heavy element
mass fractions of $> 50\%$ are possible
\citep{Baraffe2008,Mordasini2009}, and a massive core for WASP-86b can
be expected given the high metallicity of its host star. However, the
estimated value of a core mass of 210\me implies very large accretion
rates and substantial planet migration.  Therefore\textbf{,} WASP-86b
could have formed far out in the disc and consequently migrated most
probably via Type II migration \citep{Mordasini2009,Mordasini2012,
  Mordasini2014}. This scenario could explain the circular orbit of
WASP-86b \citep{Dunhill2015}, as well as the growth of a very massive
core given the larger available mass reservoir as the planet migrates
within the disk \citep{Mordasini2014}.

\subsection{Planetary envelopes}
Giant planets are expected to form far out in the protoplanetary disc
and then migrate inward. Massive cores can form at larger distances in
the protoplanetary disc, but even if the material is available in the
disc the planetesimal accretion rate must exceed the gas accretion
rate which seems unlikely for massive cores.  Additionally, WASP-86b
seems to lack a massive hydrogen-helium envelope. To better understand
the peculiarity of WASP-86b we show in Figure \ref{mass-radius} the
mass-radius parameter space for all planets as in Figure
\ref{mass-density}. WASP-86b is at the edge of the forbidden zone at
the right-bottom corner of the diagram where planets can not be found
simply because they would be too massive for their host star planetary
disc to form. In Figure \ref{mass-radius}, we also plot, as reference,
models of planetary interior composition for planets with pure ice and
50\% water -- 50\% rock from \citet{Fortney2007}; and pure water
models from \citet{Zeng2013}.  Any gas giant planet is expected to
have a core mass of about 10-15 $\me$
\citep{Alibert2005a,Mordasini2009} under the core accretion scenario
(although the exact value is still unclear \citealt{Miller2011}). Once
this core mass is reached then `run-away' gas accretion takes
place\textbf{,} and indeed every planet with \mpl$\ge 100 \me$ has an
envelope.  Figure \ref{mass-radius} shows that compared to other gas
giants of the same mass WASP-86b has a much smaller H/He envelope.

\subsection{Our formation scenarios}
\subsubsection{WASP-86b}
To reconcile all the above mentioned characteristics of WASP-86b we
propose a different formation scenario whereby the planet has formed
and migrated inward and, after the disc dispersal, the planet has
undergone a major impact with another body in the system.  Numerical
simulations show that major impacts of two super-Earth like planets,
or the merger of more massive giant planets, can lead to complete
coalescence of the two bodies \citep{Liu2015}. Moreover,
\citet{Petrovich2015} show that in unstable systems, planets at short
semi-major axes ($<0.5$AU) are more likely to collide rather than have
their orbits excited to high eccentricities and inclinations.

A similar scenario has been proposed to explain other compact planets
such as HAT-P-2b \citep{Baraffe2008}, HD~149026b \citep{Ikoma2006} and
HATS-17b \citep{Brahm2016}. In such an event the original planet can
be stripped of the majority if not all of its envelope
\citep{Ida2004,Liu2015} which could possibly explain the lack of a
large gaseous envelope for WASP-86b.  Interestingly, given the young
age of WASP-86 (about 1~Gyr), and its circular orbit, such a major
impact might just have taken place in an event similar to the late
heavy bombardment in the Solar system, which is thought to have
happened $\sim$600Myr after formation \citep{Pfalzner2015}.  Thus the
WASP-86 system could provide a window into planet formation and
dynamics at an age that was crucial for the development of our Solar
system \citep{Gomes2005}. Observational signatures of giant impacts
dissipate on timescales that are much too short to be observable in
the WASP-86 system \citep{Jackson2014}, thus an abnormally high giant
planet density might represent one of the only indirect ways in which
we could deduce the existence of a giant impact.  This hypothesis is
also supported by recent models of planet formation via population
synthesis \citep{Mordasini2012,Mordasini2014} which show in their
synthetic mass-radius relation that planets with densities as high as
that of WASP-86b, and a heavy elements enrichment of 80\%, have masses
below 40\me\ and radii smaller than 5.2\re.

\emph{\bf {Caveat for our scenario.}}  It must be noted that in the
case where the presence of the third body in the WASP-86 system is
confirmed and is a low-mass stellar companion, its light could be
diluting the transit light curve making the planet radius to appear
smaller.  We can however reject from our analysis of the spectra a
high-mass star.  High-angular resolution imaging, like Lucky Imaging
we have applied for, will enable us to determine if such dilution
exists and thus confirm the dense nature of the planet.

\subsubsection{WASP-102b}
At the opposite end of the spectrum from WASP-86b, WASP-102b is a
sub-Jupiter gas giant planet with twice the mass of Saturn that shows
a moderate radius anomaly as defined by \citet{Laughlin2011}. With a
metallicity of \feh\ = $+$0.09 dex, WASP-102b is not expected to
possess a large core. However, with a radius of 1.259\rj, WASP-102b is
$\sim$15\% larger than predicted for a coreless model of a planet
orbiting at 0.045 AU from the Sun for the age of 4.5~Gyr
\citep{Fortney2007,Baraffe2008}. Figures \ref{mass-density} \&
\ref{mass-radius} show the mass--density and mass--radius plots for
all planets as described above. WASP-102b (highlighted in blue)
belongs to the class of moderately irradiated systems showing larger
than model--predicted radii. Figure \ref{mass-radius} shows that the
maximum extent of the radius anomaly is observed for planets with
masses between $\sim$0.3~\mj~($\sim100$~\me, i.e. Saturn mass) and
0.7~\mj~(223~\me).

The planetary radius depends on multiple physical properties such as
the age, the irradiation flux, the planet's mass, the atmospheric
composition, the presence of heavy elements in the envelope or in the
core, the atmospheric circulation, and also on any source generating
extra heating in the planetary interior. Because of WASP-102b's short
orbital period and circular orbit, the amount of irradiation received
from its host star is probably the most important cause of WASP-102b's
large radius (assuming an age of 5~Gyrs).  The planet equilibrium
temperature, estimated assuming zero albedo and efficient energy
redistribution between the planet's day-- and night--sides, is
T$_{\rm{eq}}$ = 1705 $\pm$ 32 K which is close to the temperature
threshold of 2000 K, below which Ohmic heating seems to be less
important \citep{Perna2012}.  Different possible mechanisms can play a
significant role in determining WASP-102b's radius anomaly, namely
tidal heating due to unseen companions pumping up the eccentricity
\citep{Bodenheimer2001,Bodenheimer2003}, kinetic heating due to the
breaking of atmospheric waves \citep{Guillot2002}, enhanced
atmospheric opacity \citep{Burrows2007}, and semi-convection
\citep{Chabrier2007}. All the above, however, can not explain the
entirety of the observed radii (\citealt{Fortney2010},
\citealt{Leconte2010}).  Because of their radius degeneracy close-in
planets, in particular at sub-Jupiter masses (0.3 - 0.7\mj), represent
a challenge for theoretical models reproducing their radii and thus
the radius anomalies currently remains an unresolved problem in the
field of exoplanets.  More systems such as WASP-102b can help improve
our understanding of planetary radii by means of the comparison among
planets of similar mass and completely different ages and irradiation
environments.

\subsection{Tidal spin-up of WASP-102}

A possible explanation for the discrepancy between the age of WASP-102
estimated via gyrochonology and that obtained using stellar
evolutionary models (see \S \ref{mass-age}) could be tidal
interactions between the star and the planet.  If the stellar rotation
period is longer than the planetary orbital period, significant
orbital angular momentum is transferred from the orbit of a planet to
the rotation of the host star (“tidal spin-up”) via tides, yielding a
star that is rotating faster than an isolated star of the same age.

Age estimation via gyrochonology assumes a natural rotational
evolution for the star, free from any external influence. This is not
always the case; in both binary star systems and hot Jupiter
exoplanetary systems, tidal torques between bodies in close proximity
can potentially overwhelm the natural spin-down that results from
magnetic braking.  If this is the case for WASP-102 then the stellar
gyrochronological age will be underestimating its true age
\citep{Zahn75,Zahn1977,Goodman-Lackner2009,Ogilvie2014,Damiani-Lanza2015}.
Under this scenario, tidal heating would also partially explain the
large radius of WASP-102b. However, WASP-102b would remain a larger
than model-expected planet for its mass, orbital distance and age.

Higher stellar rotation rates (measured either directly, or by means
of \vsini) in systems with hot Jupiters have been proposed as
indications of a tidal influence of exoplanets on their host star
\citep{Husnoo2012}. Under the assumption that the star has been spun
up by the presence of the planet, then from the increase in stellar
rotation we would also expect an increase in stellar activity because
rotation is a major driver of activity
\citep{Poppenhaeger2014}. However, we do not detect any indication of
stellar activity in our spectra (Ca II H\&K) nor in the analysis of
the WASP data.

Recent studies of the discrepancy between stellar ages estimated via
gyrochronology and those estimated using stellar models highlight an
existing bias towards older ages when using isochronal analysis owing
to the uneven spacing of data in isochrones near the ZAMS
\citep{Brown2014,Soderblom2010,Barnes2007,Saffe2005}.  Recently,
\citet{Maxted2015b} explored this discrepancy for 28 exoplanet host
stars and found strong evidence showing that, for half of the stars in
their sample, gyrochronological ages of planet host stars are
significantly lower than the isochronal ages. This is also the case
for WASP-102b. Given WASP-102b's short orbital period and relatively
large mass, it is possible that tidal interactions have spun up the
planet's host star. However, given that \citet{Maxted2015b} show that
the age discrepancy is not always good evidence for tidal
interactions, more observational evidence would be necessary in order
to confirm tidal spin-up of the star.

\section{Conclusion}

In this paper we have reported the discovery of two new gas giant
planets, WASP-86b and WASP-102b, from the SuperWASP survey.  New
discoveries from ground-based transit surveys show that the currently
known population of gas giant planets is by no means exhaustive. More
peculiar systems such as WASP-86b and many others, including WASP-127b
\citep{Lam2016}, HATS-18b \citep{Panev2016}, and EPIC212521166 b
\citep{Osborn2016}, just to mention a few, continue to provide
invaluable observable constraints for theoretical models of formation
and evolution of planetary systems.

\begin{acknowledgements}

  The SuperWASP Consortium consists of astronomers primarily from
  University of Warwick, Queens University Belfast, St Andrews, Keele,
  Leicester, The Open University, Isaac Newton Group La Palma and
  Instituto de Astrof\'isica de Canarias. The SuperWASP-N camera is
  hosted by the Issac Newton Group on La Palma and WASPSouth is hosted
  by SAAO. We are grateful for their support and assistance. Funding
  for WASP comes from consortium universities and from the UK's
  Science and Technology Facilities Council. The research leading to
  these results has received funding from the European Community's
  Seventh Framework Programmes (FP7/2007-2013 and FP7/2013-2016) under
  grant agreement number RG226604 and 312430 (OPTICON),
  respectively. D.J.A. acknowledges funding from the European Union Seventh Framework programme (FP7/2007-2013) under grant agreement No. 313014 (ETAEARTH). D.J.A.B acknowledges funding from the UKSA and the University of Warwick.
  Based on observations made with the CORALIE Echelle spectrograph
  mounted on the Swiss telescope at ESO La Silla in Chile.  TRAPPIST
  is funded by the Belgian Fund for Scientific Research (Fond National
  de la Recherche Scientifique, FNRS) under the grant FRFC
  2.5.594.09.F, with the participation of the Swiss National Science
  Fundation (SNF).  
  Based upon observations carried out at the Observatorio Astron\'omico Nacional on the Sierra San Pedro M\'artir (OAN-SPM), Baja California, M\'exico.
  Used Simbad, Vizier, exoplanet.eu
\end{acknowledgements}

\bibliographystyle{aa}
\bibliography{FF}

\end{document}